\documentstyle[aps,preprint,epsf,axodraw]{revtex}      
\newcounter{eqnaux}
\newcounter{eqntmp}
\tightenlines
\begin{document}
\title{Relativistic Quark Spin Coupling Effects in the
Correlations Between Nucleon Electroweak Properties}
\author{E.F. Suisso$^a$, W.R.B. de Ara\'ujo$^b$, 
T. Frederico$^a$, M. Beyer$^c$, and
  H.J. Weber$^d$}
\address{$^a$ Dep. de F\'\i sica, Instituto Tecnol\'ogico de Aeron\'autica,
Centro T\'ecnico Aeroespacial, \\
12.228-900 S\~ao Jos\'e dos Campos, S\~ao Paulo, Brazil.}
\address{$^b$Laborat\'orio do Acelerador Linear, Instituto de F\'\i sica
da USP\\ C.P. 663118, CEP 05315-970, S\~ao Paulo, Brazil}
\address{$^c$ Fachbereich Physik, 
Universit\"at Rostock, 18051 Rostock, Germany}
\address{$^d$ Dept. of Physics, University of Virginia, 
Charlottesville, VA 22901, U.S.A.}
\maketitle
\begin{abstract}
  We investigate the effect of different relativistic spin couplings
  of constituent quarks on nucleon electroweak properties. Within each
  quark spin coupling scheme the correlations between static
  electroweak observables are found to be independent of the
  particular shape of the momentum part of the nucleon light-front
  wave function.  The neutron charge form factor is very sensitive to
  different choices of spin coupling schemes once the magnetic moment
  is fitted to the experimental value. However, it is found rather
  insensitive to the details of the momentum part of the three-quark
  wave function model. 

\end{abstract}

\vspace{3ex}

\section{Introduction}

In a previous work\cite{asfbw}, we have studied nucleon
electromagnetic form factors using different forms of relativistic
spin couplings between the constituent quarks forming the nucleon. We
have used an effective Lagrangian to describe the quark spin coupling
to the nucleon keeping close contact with covariant field theory. We
have performed a three-dimensional reduction of the amplitude for the
(virtual) photon absorption by the nucleon to the null-plane,
$x^+=x^0+x^3=0$, (see, e.g., Ref.~\cite{karmanov}).  After the
three-dimensional reduction the momentum part of the nucleon
light-front wave function was introduced into the two-loop momentum
integrations that define the matrix elements of the electromagnetic
current.

In Ref.~\cite{asfbw} we have tested different spin couplings for the
nucleon in a calculation of nucleon electromagnetic form factors and
found that the neutron charge form factor in particular leads to
constraints of the quark spin coupling.  The comparison with the
neutron data below momentum transfer of 1 GeV/c suggests that the
scalar pair is preferred in the relativistic quark spin coupling of
the nucleon. That study was performed assuming the same Gaussian wave
function for both the mixed scalar and gradient quark pair couplings.

Presently, while extending this investigation to other form factors we
additionally introduce a power law behavior for the momentum part of
the light-front wave function.  The purpose is to investigate whether
the neutron charge form factor is still reproduced with a scalar quark
pair coupling while relaxing the form of the momentum part of the
light-front wave function.  This is indeed the case for both forms
(Gaussian and power-law) once the magnetic moment of the neutron is
fitted to its experimental value.  Moreover, for a given quark spin
coupling scheme and independent of the shape of the light-front wave
function a model independent relation between the neutron charge
radius and its magnetic moment can be recognized.  We also present
results on the nucleon axial vector form factor and on correlations
between the static electroweak observables for different spin
couplings and wave functions. In the context of the Bakamjian-Thomas
(BT) quark spin coupling scheme it was shown that the axial vector
coupling constant, the proton magnetic moment, and the radius are
correlated by model independent relations\cite{bsch,brodsky}.  We
point out that the high momentum transfer calculation of the nucleon
electromagnetic form factors with that model were first done in
Ref.~\cite{fra96}. We show that the different quark spin coupling
schemes retain the model independent correlations found.  However, the
relations involving the axial vector coupling constant obtained with a
spin coupling scheme from an effective Lagrangian differ from those
derived within the Bakamjian-Thomas construction \cite{bsch}.

The effective Lagrangian for the N-q coupling is written
as\cite{asfbw},
\begin{eqnarray}
{\cal{L}}_{\mathrm{N-3q}}=\alpha m_N\epsilon^{lmn}
\overline{\Psi}_{(l)}
i\tau _2\gamma^5\Psi_{(m)}^C\overline{\Psi} _{(n)}\Psi _N
+(1-\alpha) 
\epsilon^{lmn}
\overline{\Psi}_{(l)}
i\tau _2\gamma_\mu \gamma ^5\Psi_{(m)}^C
\overline{\Psi} _{(n)}i\partial^\mu\Psi _N
+ H.C.
\label{lag}
\end{eqnarray}
where $\tau _2$ is the isospin matrix, the color indices are
$\{l,m,n\}$ and $\epsilon^{lmn}$ is the totally antisymmetric symbol.
The conjugate quark field is $\Psi^C=C \overline{\Psi}^\top $, where
$C=i \gamma^2\gamma^0$ is the charge conjugation matrix; $\alpha$ is a
parameter to vary the relative magnitude of the spin couplings, and
$m_N$ is the nucleon mass.

The macroscopic matrix elements of the nucleon electromagnetic current
$J^+_N(q^2)$ in the Breit-frame and in the light-front spinor basis is
given by:
\begin{eqnarray}
\langle s'|J^+_N(q^2)|s\rangle &=&\bar{u}(p',s')
\left( F_{1N}(q^2)\gamma^++ i \frac{\sigma^{+\mu}q_\mu}{2 m_N}F_{2N}(q^2)
\right) {u}(p,s)
\nonumber \\
&=& \frac{p^+}{m_N}
\langle s'| F_{1N}(q^2)- i \frac{F_{2N}(q^2)}{2 m_N}
\vec q_\perp\cdot (\vec n \times \vec \sigma )| s \rangle \ ,
\label{jp}
\end{eqnarray}
where $F_{1N}$ and $F_{2N}$ are the Dirac and Pauli form factors,
respectively, while $\vec n$ is the unit vector along the
$z$-direction.  The Breit-frame momenta are $q=(0,\vec q_\perp,0)$,
such that $(q^+=q^0+q^3=0)$ and $\vec q_\perp=(q^1,q^2)$;
$p=(\sqrt{q^2_\perp/4+m^2_N},-\vec q_\perp/2,0)$ and
$p'=(\sqrt{q^2_\perp/4+m^2_N},\vec q_\perp/2,0)$.

The Sachs form factors are defined by:
\begin{eqnarray}
G_{EN}(q^2)&=& F_{1N}(q^2)+\frac{q^2}{4m_N^2}F_{2N}(q^2) \ ,
\nonumber \\
G_{MN}(q^2)&=& F_{1N}(q^2)+F_{2N}(q^2) \ .
\end{eqnarray}
The magnetic moment is $\mu_N=G_{MN}(0)$ and the mean squared
radius is $r^2_N=6\frac{dG_{EN}(q^2)}{dq^2}|_{q^2=0}$.

The non-vanishing part of the macroscopic matrix elements of 
the nucleon weak isovector axial vector current $A^+_N(q^2)$ 
in the Breit-frame 
with $q^+=0$   in the light-front spinor basis is given by:
\begin{eqnarray}
\langle s'|A^+_{N}(q^2)|s\rangle &=&\bar{u}(p',s')
\left( G_{A}(q^2)\gamma^+\gamma^5 \frac{\vec \tau}{2}  
\right) {u}(p,s)
\nonumber \\
&=& \frac{p^+}{m_N}G_A(q^2)
\langle s'| \frac{\vec \tau}{2} \sigma_z | s \rangle \ ,
\label{ap}
\end{eqnarray}
where $G_A$ is the  weak isovector axial vector form factor and $g_A=G_A(0)$ is
the axial vector coupling constant.

The light-front spinors are:
\begin{eqnarray}
u(p,s) =\frac{\rlap\slash p+ m}{2\sqrt{p^+m}}\gamma^+
\gamma^0 \left(\begin{array}{c}
\chi^{\rm Pauli}_{s} \\0\end{array}
\right)
\ .
\label{lf}
\end{eqnarray}
The Dirac spinor of the instant form
\begin{eqnarray}
u_{D}(p,s)=\frac{\rlap\slash p
+m}{\sqrt{2 m(p^0+m) }}
\left(\begin{array}{c}
\chi^{\rm Pauli}_{s} \\
0
\end{array}
\right)  \
\label{dirac}
\end{eqnarray}
carries the subscript $D$.
The Melosh rotation is the unitary transformation between the
light-front and instant form spinors that is given by:
\begin{eqnarray}
\left[R_{M}(p)\right]_{s's}
\ =\langle s'|\frac{p^++m-i\vec \sigma . (\vec n\times \vec p)}{
\sqrt{(p^++m)^2+p^2_\perp}}| s \rangle
\ =\overline{u}_D(p,s')u(p,s)
\ ,
\label{mel}
\end{eqnarray}
where $\vec n$ the unit vector along the z-direction.

In section II, the general form of the microscopic matrix elements of
the nucleon electroweak current are discussed. The detailed form of
the electromagnetic current is derived and the light-front wave
function is introduced in the computation of the form factors. Also,
the matrix element of the weak isovector axial vector current of the
nucleon are derived from the effective Lagrangian. In section III, the
physics of the different spin coupling schemes are discussed in
comparison with the widely used Bakamjian-Thomas framework. In section
IV, the numerical results of the static electroweak observables and of
the form factors are presented. The model independence within each
spin coupling scheme is demonstrated for the correlation between the
static nucleon electroweak observables.  In section V, we give the
summary and conclusion.

\section{Nucleon electroweak current}

The microscopic matrix elements of the nucleon electromagnetic and
weak isovector axial vector currents are constructed from the
effective Lagrangian given in Eq.(\ref{lag}). The current matrix
elements are evaluated in impulse approximation. The complete
antisymmetrization of the quark states implies four topologically
distinct diagrams depicted in Figure 1.  The two-loop triangle
diagrams of Figure 1 represent the impulse approximation for the
evaluation of the baryon form factors in light-front dynamics. We
calculate the matrix elements of the currents via coupling to the
third quark due to the symmetrization of the microscopic matrix
element after factorizing the color degree of freedom.  The
electromagnetic quark current operator is $\overline \Psi \hat
Q_q\gamma^\mu \Psi$, with $\hat Q_q$ the charge operator, and the weak
isovector axial vector current one is $\overline \Psi \frac{\vec
  \tau}{2}\gamma^\mu\gamma^5 \Psi$.

In detail, Figure 1a represents the nucleon spin-space operators
$J^+_{aN}$ and $A^+_{aN}$. In these cases the elementary operators act
on quark 3 while 1 and 2 compose the coupled spectator quark pair of
Eq.~(\ref{lag}) for the initial and final nucleons alike. In Figure
1b, the coupled quark pair of the initial nucleon is (13) whereas it
is (12) in the final nucleon. The operators $J^+_{bN}$ and $A^+_{bN}$
represented by Figure 1b are multiplied by a factor of 4.  A factor 2
comes from the exchange of quarks 1 and 2 and another factor 2 comes
{}from the invariance under exchanging the pairs in the initial and
final nucleons that is a consequence of time reversal and parity
transformation properties. The operators $J^+_{cN}$ and $A^+_{cN}$ are
represented by Figure 1c, where the initial coupled pair quark is (13)
and the final coupled pair is (23).  This operator is multiplied by a
factor of 2 because quarks 1 and 2 can be exchanged.  Finally, the
process shown in Figure 1d does not contribute to the nucleon axial
vector current because of the isoscalar quark pair as given by the
Lagrangian of Eq.(\ref{lag}). However this diagram is non vanishing
for the electromagnetic current and denoted by $J^+_{dN}$. It
corresponds to the process in which the photon is absorbed by the
coupled quark pair (13) while 2 is the spectator. In this case, two
diagrams are possible by the exchange of quarks 1 and 2 giving rise to
a factor of 2.

The microscopic operator of the nucleon electromagnetic current 
is given by the sum of four terms:
\begin{eqnarray}
J^+_N(q^2)=J^+_{aN}(q^2)+4J^+_{bN}(q^2)+2J^+_{cN}(q^2)+2J^+_{dN}(q^2)
.
\label{mjp}
\end{eqnarray}
The weak isovector axial vector current has contribution from three terms:
\begin{eqnarray}
A^+_N(q^2)=A^+_{aN}(q^2)+4A^+_{bN}(q^2)+2A^+_{cN}(q^2)
\ .
\label{ajp}
\end{eqnarray}
The term $A^+_{dN}(q^2)$ vanishes because of isospin properties.

\subsection{Derivation of the Electromagnetic Current Matrix Elements}

The nucleon current operators $J^+_{\beta N}$, $\beta=a,b,c,d$ and
$A^+_{\gamma N}$, $\gamma=a,b,c$ of Eqs.~(\ref{mjp}) and (\ref{ajp})
are constructed directly from the Feynman diagrams of Figure 1. The
electromagnetic current $J^+_{N}$ receives contributions from each
amplitude represented by the Feynman two-loop  triangle diagrams of
Figures 1a to 1d, which we repeat here\cite{asfbw}:
\begin{eqnarray}
\langle s'|J^+_{a N}(q^2)|s\rangle &=& -
\langle N|\hat Q_q| N\rangle {\mathrm{Tr}}[i \tau_2(-i )\tau_2]
\int \frac{d^4k_1d^4k_2}{(2\pi)^8}\Lambda(k_i,p^{\prime})\Lambda(k_i,p)
\bar u(p',s')S(k'_3)\gamma^+
\nonumber \\
&\times& S(k_3)u(p,s)
{\mathrm{Tr}}\left[S(k_2)\left(\alpha m_N+(1-\alpha)\rlap\slash p\right)
\gamma^5 S_c(k_1)\gamma^5\left(\alpha m_N+(1-\alpha)\rlap\slash p'\right)
\right]
\ ,  \label{j+a}
\end{eqnarray}
with $\displaystyle S(p)=\frac{1}{\rlap\slash p-m+i  \epsilon} \, ,$
and $\displaystyle S_c(p)=\left[\gamma^0\gamma^2
\frac{1}{\rlap\slash p-m+i  \epsilon}\gamma^0\gamma^2\right]^\top \, $.
Here $m$ is the constituent quark mass and $k'_3=k_3+Q \, $,
and $\langle N|\hat Q_q| N\rangle$ is the isospin matrix.
The function $\displaystyle \Lambda(k_i,p)$ is chosen to
introduce the momentum part of the three-quark light-front wave function,
after the integrations over $k^-$ are performed. 
The contribution to the electromagnetic current represented by Figure
1b is given by:
\begin{eqnarray}
\langle s'|J^+_{b N}(q^2)|s\rangle &=& - \langle N|\hat Q_q| N\rangle
\int \frac{d^4k_1d^4k_2}{(2\pi)^8}\Lambda(k_i,p^{\prime})
\Lambda(k_i,p)
\bar u(p',s')S(k'_3)\gamma^+S(k_3)
\nonumber \\
&\times& \left(\alpha m_N+(1-\alpha)\rlap\slash p\right)\gamma^5
S_c(k_1)\gamma^5\left(\alpha m_N+(1-\alpha)\rlap\slash p'\right)
 S(k_2)u(p,s)
\ .  \label{j+b}
\end{eqnarray}
The contribution to the electromagnetic current represented by
 Figure 1c is given by:
\begin{eqnarray}
\langle s'|J^+_{c N}(q^2)|s\rangle &=& 
\langle N|\tau_2\hat Q_q\tau_2| N\rangle
\int \frac{d^4k_1d^4k_2}{(2\pi)^8}\Lambda(k_i,p^{\prime})
\Lambda(k_i,p) \bar u(p',s')S(k_1)
\left(\alpha m_N+(1-\alpha)\rlap\slash p\right)
\nonumber \\
&\times&
\gamma^5 S_c(k_3)\gamma^+
S_c(k'_3)\gamma^5\left(\alpha m_N+(1-\alpha)\rlap\slash p'\right)
S(k_2)u(p,s)
\ .  \label{j+c}
\end{eqnarray}
The contribution to the electromagnetic current  represented by
Figure 1d is given by:
\begin{eqnarray}
\langle s'|J^+_{d N}(q^2)|s\rangle &=&-
{\mathrm{Tr}}
[\hat Q_q]\int \frac{d^4k_1d^4k_2}{(2\pi)^8}\Lambda(k_i,p^{\prime})
\Lambda(k_i,p)\bar u(p',s')S(k_2)u(p,s)
\nonumber \\
&&{\mathrm{Tr}}\left[
\gamma^5\left(\alpha m_N+(1-\alpha)\rlap\slash p'\right)S(k'_3)\gamma^+
S(k_3)\left(\alpha m_N+(1-\alpha)\rlap\slash p\right)
\gamma^5 S_c(k_1)\right]
\ .  \label{j+d}
\end{eqnarray}

The light-front coordinates are defined as $k^+=k^0+k^3\ , k^-=k^0-k^3
\ , k_\perp=(k^1,k^2).$ In each term of the nucleon current, from
$J^+_{aN}$ to $J^+_{dN}$, the Cauchy integrations over $k^-_1$ and
$k^-_2$ are performed.  That means the on-mass-shell pole of the
Feynman propagators for the spectator particles 1 and 2 of the photon
absorption process are taken into account.  In the Breit-frame with
$q^+=0$ there is a maximal suppression of light-front Z-diagrams in
$J^+$ \cite{tob92,pach99}.  Thus the components of the momentum
$k^+_1$ and $k^+_2$ are bounded such that $ 0< k^+_1 < p^+$ and
$0<k^+_2 <p^+-k^+_1$\cite{pach97}. The four-dimensional integrations
of Eqs.(\ref{j+a}) to (\ref{j+d}) are reduced to the three-dimensional
ones of the null-plane.

After the integrations over the light-front energies 
the momentum part of the wave function is introduced into
the microscopic matrix elements of the current 
by the substitution \cite{asfbw,tob92}:
\begin{eqnarray}
\frac{1}{2(2\pi)^3}
\frac{\Lambda(k_i,p)}{m^2_N-M^2_0}\rightarrow  \Psi (M^2_0)
\ .
\end{eqnarray}
To study the model dependence we choose the harmonic wave function
and a power-law form \cite{bsch,brodsky},
\setcounter{eqnaux}{\value{equation}}
\begin{equation}
\Psi_{\mathrm{HO}}=N_{\mathrm{HO}}\exp(-M^2_0/2\beta^2),\qquad 
\Psi_{\mathrm{Power}}=N_{\mathrm{Power}}(1+M^2_0/\beta^2)^{-p}
\ ,
\end{equation}
and $\beta$ is the width parameter. The free three-quark mass $M_0$ is
given below in Eq.(\ref{eqn:M0}). From perturbative QCD arguments a
power-law fall-off with $p=3.5$ is predicted \cite{brodsky}. The
relations between static electroweak observables are not sensitive to
$p$ as long as $p>2$ \cite{bsch}. We choose for our calculations
$p=3$.  Further, the same momentum wave function is chosen all N-q
couplings, for simplicity. Note, that the mixed ($\alpha=1/2$) case
could have different momentum dependencies for each spin coupling,
however, we choose the same momentum functions just to keep contact
with the BT approach.

The analytical integration of Eq.(\ref{j+a}) of the $k^-$ components of
the momenta yields:
\begin{eqnarray}
\langle s'|J^+_{a N}(q^2)|s\rangle  &=& 2p^{+2}
\langle N|\hat Q_q|N\rangle
\int \frac{d^{2} k_{1\perp} dk^{+}_1d^{2} k_{2\perp} d
k^{+}_2 }{k^+_1k^+_2k^{+\ 2}_3} \theta(p^+-k^+_1)
\theta(p^+-k^+_1-k^+_2) \nonumber \\
&&{\mathrm{Tr}}
\left[ (\rlap\slash k_2+m)\left(\alpha m_N+(1-\alpha)\rlap\slash p\right)
(\rlap\slash k_1+m)\left(\alpha m_N+(1-\alpha)\rlap\slash p'\right)\right]
\nonumber \\
&&\bar u(p',s')(\rlap\slash k'_3+m))\gamma^+(\rlap\slash k_3+m)u(p,s)
\Psi (M^{'2}_0)
\Psi (M^2_0)
 \ ,
\label{j+alf}
\end{eqnarray}
where $k^2_1=m^2$ and $k^2_2=m^2$. The free three-quark squared mass
is defined by:
\begin{equation}
M^2_0=p^+(\frac{k_{1\perp}^{2}+m^2}{k^+_1}+\frac{k_{2\perp}^{2}+m^2}{k^+_2}
+\frac{k_{3\perp}^{2}+m^2}{k^+_3})-{p^2_\perp} \ ,
\label{eqn:M0}
\end{equation}
and $M^{\prime 2}_0=M^2_0(k_3\rightarrow k'_3 \ , \vec p_\perp\rightarrow
\vec p^\prime_\perp)$.

The other terms of the nucleon current, as given by Eqs.
(\ref{j+b})-(\ref{j+d}) are also integrated over the $k^-$ momentum components
of  particles 1 and 2 following the same steps used to
obtain Eq.(\ref{j+alf}) from Eq.(\ref{j+a}):
\begin{eqnarray}
\langle s'|J^+_{b N}(q^2)|s\rangle  &=& p^{+2}
\langle N|\hat Q_q| N\rangle
\int \frac{d^{2} k_{1\perp} dk^{+}_1d^{2} k_{2\perp} d
k^{+}_2 }{
k^+_1k^+_2k^{+\ 2}_3} \theta(p^+-k^+_1)
\theta(p^+-k^+_1-k^+_2) \nonumber \\
&&\bar u(p',s')(\rlap\slash k'_3+m)\gamma^+(\rlap\slash k_3+m)
\left(\alpha m_N+(1-\alpha)\rlap\slash p\right) (\rlap\slash k_1+m)
\nonumber \\
&&\times \left(\alpha m_N+(1-\alpha)\rlap\slash p'\right)
 (\rlap\slash k_2+m)u(p,s)
\Psi (M^{'2}_0)
\Psi (M^2_0)
\ ,  \label{j+blf}
\end{eqnarray}
\begin{eqnarray}
\langle s'|J^+_{c N}(q^2)|s\rangle  &=&
p^{+2}\langle N|\tau_2\hat Q_q \tau_2| N\rangle
\int \frac{d^{2} k_{1\perp} dk^{+}_1d^{2} k_{2\perp} d
k^{+}_2 }{
k^+_1k^+_2k^{+\ 2}_3} \theta(p^+-k^+_1)
\theta(p^+-k^+_1-k^+_2) \nonumber \\
&&\bar u(p',s')(\rlap\slash k_1+m)
\left(\alpha m_N+(1-\alpha)\rlap\slash p\right)
(\rlap\slash k_3+m)\gamma^+
(\rlap\slash k'_3+m)
\nonumber \\
&&
\times \left(\alpha m_N+(1-\alpha)\rlap\slash p'\right)
 (\rlap\slash k_2+m)u(p,s)
\Psi (M^{'2}_0)
\Psi (M^2_0)
\ ,  \label{j+clf}
\end{eqnarray}
\begin{eqnarray}
\langle s'|J^+_{d N}(q^2)|s\rangle  &=& p^{+2}{\mathrm{Tr}}[\hat Q_q]
\int \frac{d^{2} k_{1\perp} dk^{+}_1d^{2} k_{2\perp} d
k^{+}_2 }{
k^+_1k^+_2k^{+\ 2}_3} \theta(p^+-k^+_1)
\theta(p^+-k^+_1-k^+_2)
\nonumber \\
&&{\mathrm{Tr}}\left[
\left(\alpha m_N+(1-\alpha)\rlap\slash p'\right)(\rlap\slash k'_3+m)
\gamma^+ (\rlap\slash k_3+m)\left(\alpha m_N+(1-\alpha)\rlap\slash p\right)
(\rlap\slash k_1+m)\right]
\nonumber \\
&&\bar u(p',s')(\rlap\slash k_2+m)u(p,s)
\Psi (M^{'2}_0)
\Psi (M^2_0)
\ .  \label{j+dlf}
\end{eqnarray}
 The normalization is chosen such that the proton
charge is unity.

\subsection{Derivation of the Axial Vector Current Matrix Elements}

The weak isovector axial vector current $A^+_{N}$ receives
contributions from each amplitude represented by the Feynman two-loop
triangle diagrams of Figures 1a to 1c:
\begin{eqnarray}
\langle s'|A^+_{a N}(q^2)|s\rangle &=& -
\langle N|\frac{\vec \tau}{2}| N\rangle 
{\mathrm{Tr}}[i \tau_2(-i )\tau_2]
\int \frac{d^4k_1d^4k_2}{(2\pi)^8}\Lambda(k_i,p^{\prime})\Lambda(k_i,p)
\bar u(p',s')S(k'_3)\gamma^+\gamma^5
\nonumber \\
&\times& S(k_3)u(p,s)
{\mathrm{Tr}}\left[S(k_2)\left(\alpha m_N+(1-\alpha)\rlap\slash p\right)
\gamma^5 S_c(k_1)\gamma^5\left(\alpha m_N+(1-\alpha)\rlap\slash p'\right)
\right]
\ .  \label{a+a}
\end{eqnarray}
The contribution to the axial vector current  
represented by Figure 1b is given by:
\begin{eqnarray}
\langle s'|A^+_{b N}(q^2)|s\rangle &=& - \langle N|\frac{\vec \tau}{2}
| N\rangle
\int \frac{d^4k_1d^4k_2}{(2\pi)^8}\Lambda(k_i,p^{\prime})
\Lambda(k_i,p)
\bar u(p',s')S(k'_3)\gamma^+\gamma^5S(k_3)
\nonumber \\
&\times& \left(\alpha m_N+(1-\alpha)\rlap\slash p\right)\gamma^5
S_c(k_1)\gamma^5\left(\alpha m_N+(1-\alpha)\rlap\slash p'\right)
 S(k_2)u(p,s)
\ .  \label{a+b}
\end{eqnarray}
The contribution to the axial vector current represented by
 Figure 1c is given by:
\begin{eqnarray}
\langle s'|A^+_{c N}(q^2)|s\rangle &=& 
\langle N|\tau_2\frac{\vec \tau}{2}\tau_2| N\rangle
\int \frac{d^4k_1d^4k_2}{(2\pi)^8}\Lambda(k_i,p^{\prime})
\Lambda(k_i,p) \bar u(p',s')S(k_1)
\left(\alpha m_N+(1-\alpha)\rlap\slash p\right)
\nonumber \\
&\times&
\gamma^5 S_c(k_3)\gamma^+\gamma^5
S_c(k'_3)\gamma^5\left(\alpha m_N+(1-\alpha)\rlap\slash p'\right)
S(k_2)u(p,s)
\ .  \label{a+c}
\end{eqnarray}
The contribution to the axial vector current represented by Figure 1d
vanishes because of the isoscalar nature of the coupled quark pair.

In each term of the nucleon axial vector current, from $A^+_{aN}$ to
$A^+_{cN}$, the Cauchy integrations over $k^-_1$ and $k^-_2$ are
performed as discussed in the previous section for the electromagnetic
current. The spectator particles are on their mass-shell after the
integrations on the $k^-$ momentum in Eqs. (\ref{a+a}) to (\ref{a+c}).
The numerators of the Dirac propagators of quark 3 on which the
axial operator $\gamma^+\gamma^5$ acts have the momenta $k'_3$
and $k_3$ on the $ k^-$-shell because $(\gamma^+)^2=0$.  The
components of the momentum $k^+_1$ and $k^+_2$ are bounded by $0<
k^+_1 < p^+$ and $0<k^+_2 <p^+-k^+_1$\cite{pach97}. The
four-dimensional integrations of Eqs.(\ref{a+a}) to (\ref{a+c}) are
reduced to the three dimensions of the null-plane.

The analytical integration of Eq.(\ref{a+a}) of the $k^-$ components of
the momenta yields:
\begin{eqnarray}
\langle s'|A^+_{a N}(q^2)|s\rangle  &=& 2p^{+2}
\langle N|\frac{\vec\tau}{2}| N\rangle 
\int \frac{d^{2} k_{1\perp} dk^{+}_1d^{2} k_{2\perp} d
k^{+}_2 }{
k^+_1k^+_2k^{+\ 2}_3} \theta(p^+-k^+_1)
\theta(p^+-k^+_1-k^+_2) \nonumber \\
&&{\mathrm{Tr}}
\left[ (\rlap\slash k_2+m)\left(\alpha m_N+(1-\alpha)\rlap\slash p\right)
(\rlap\slash k_1+m)\left(\alpha m_N+(1-\alpha)\rlap\slash p'\right)\right]
\nonumber \\
&&\bar u(p',s')(\rlap\slash k'_3+m))\gamma^+\gamma^5
(\rlap\slash k_3+m)u(p,s)
\Psi (M^{'2}_0)
\Psi (M^2_0)
 \ ,
\label{a+alf}
\end{eqnarray}
and $k^2_1=m^2$ and $k^2_2=m^2$.

The integrations in the light-front energies in  Eqs. (\ref{a+b}) and 
(\ref{a+c}) lead to:
\begin{eqnarray}
\langle s'|A^+_{b N}(q^2)|s\rangle  &=& p^{+2}
\langle N|\frac{\vec\tau}{2}| N\rangle 
\int \frac{d^{2} k_{1\perp} dk^{+}_1d^{2} k_{2\perp} d
k^{+}_2 }{
k^+_1k^+_2k^{+\ 2}_3} \theta(p^+-k^+_1)
\theta(p^+-k^+_1-k^+_2) \nonumber \\
&&\bar u(p',s')(\rlap\slash k'_3+m)\gamma^+\gamma^5(\rlap\slash k_3+m)
\left(\alpha m_N+(1-\alpha)\rlap\slash p\right) (\rlap\slash k_1+m)
\nonumber \\
&&\times \left(\alpha m_N+(1-\alpha)\rlap\slash p'\right)
 (\rlap\slash k_2+m)u(p,s)
\Psi (M^{'2}_0)
\Psi (M^2_0)
\ ,  \label{a+blf}
\end{eqnarray}
\begin{eqnarray}
\langle s'|A^+_{c N}(q^2)|s\rangle  &=&
p^{+2}\langle N|\tau_2\frac{\vec \tau}{2} \tau_2| N\rangle
\int \frac{d^{2} k_{1\perp} dk^{+}_1d^{2} k_{2\perp} d
k^{+}_2 }{
k^+_1k^+_2k^{+\ 2}_3} \theta(p^+-k^+_1)
\theta(p^+-k^+_1-k^+_2) \nonumber \\
&&\bar u(p',s')(\rlap\slash k_1+m)
\left(\alpha m_N+(1-\alpha)\rlap\slash p\right)
(\rlap\slash k_3+m)\gamma^+\gamma^5
(\rlap\slash k'_3+m)
\nonumber \\
&&
\times \left(\alpha m_N+(1-\alpha)\rlap\slash p'\right)
 (\rlap\slash k_2+m)u(p,s)
\Psi (M^{'2}_0)
\Psi (M^2_0)
\ ,  \label{a+clf}
\end{eqnarray}

\section{Discussion of Spin Coupling Schemes}

The physical meaning of the effective Lagrangian for the quark spin
coupling emerges if one performs a kinematical light-front boost of
the matrix elements of the spin operators between quark states on one
hand and quark-nucleon states related to the initial and final
nucleons with their respective rest frames on the other hand. This has
been suggested in Ref.\cite{araujo99} and also discussed in
Ref.\cite{asfbw}. The effective Lagrangian of Eq.(\ref{lag}) contains
the spin-flavor invariants of the nucleon with quark pair spin zero
($\alpha =1$) and spin one ($\alpha=0$) that are 2 of a basis of 8
such states given in detail in Ref.~\cite{BKW98}.  The nucleon spin
invariant that is widely used and tested in form factor calculations
uses the ones chosen here but contain the additional projector
$\rlap\slash p +M_0$ onto large Dirac components, a characteristic
feature of the Bakamjian-Thomas (BT) spin coupling scheme \cite{BT53}.
The spin-flavor invariant of the effective Lagrangian Eq.~(\ref{lag})
with $\alpha=1/2$ resembles the BT spin coupling scheme but is not
equivalent to it, i.e., the Melosh rotations have their arguments
defined in the nucleon rest frame with individual '+' momentum
constrained by the total nucleon $p^+$. The BT construction have the
Melosh spin rotation with the individual '+' momentum constrained by
the free three-quark mass $M_0$.  That differs from the above
Lagrangian as explicitly shown in Ref.\cite{araujo99}. Moreover, in
the pointlike nucleon limit, the weak isovector axial vector coupling
constant represents a situation in which the difference between BT and
effective Lagrangian spin coupling schemes is maximized as we will
discuss at the end of this section.

The Melosh rotations appear in the equations for the vector and axial
vector current from the residues of the triangle Feynman diagram,
which are evaluated at the on-$k^-$-shell poles of the spectator
particles, and each of the numerators of the Dirac propagator are
on-$k^-$-shell. In particular, the numerator of quark 3 comes to be
on-$k^-$-shell because $(\gamma^+)^2=0$.  Consequently, the numerators
of the fermion propagators are substituted by the positive energy
spinor projector, written in terms of light-front spinors. We use that
the Wigner rotation is unity for kinematical Lorentz transformations
to calculate the spin matrix elements of the nucleon current
corresponding to the respective rest-frames of the initial or final
nucleon.  A typical matrix element of the spin coupling coefficient
for $\alpha=1$ appearing in the evaluation of $J^+$ as well as in
$A^+$, when calculated in the nucleon rest frame, is given by:
\begin{equation}
\chi (s_1,s_2,s_3;s_N)=\overline{u}_{1}
\gamma ^5 u_{2}^C\;
\overline{u}_{3}u_{N} \ ,
\label{nuc}
\end{equation}
where  $u_i=u(k_i,s_i)$ is the light-front spinor for the
$i$-th quark.

The matrix element of the pair coupled to spin zero in Eq.~(\ref{nuc})
is evaluated in the rest frame of the pair (c.m.) reached 
by a kinematical light-front boost from the nucleon rest frame. 
The Wigner rotation is unity for such a Lorentz transformation
consequently
 (viz. ${u}_{\rm c.m.}(\vec k^{{\rm c.m.}},s) = {u}(\vec
k^{{\rm c.m.}},s)$):
\begin{eqnarray}
I(s_1,s_2,0)&=&\overline{u}(\vec k_1,s_1)
\gamma ^5u^C(\vec k_2,s_2)\nonumber\\
&=& \overline{u}(\vec k^{{\rm c.m.}}_1,s_1)
\gamma^5 u^C(\vec k^{{\rm c.m.}}_2,s_2)
\ , \label{eqn:i}
\end{eqnarray}
where the particle momenta  in the pair (12) rest frame
are $\vec k^{{\rm c.m.}}=(k^{+{\rm c.m.}},{\vec k}_\perp^{{\rm c.m.}})$ 
obtained from $k^{{\rm (c.m.)}\mu}=(\Lambda k)^\mu$.
The operator $\Lambda$ is the  kinematical light-front transformation 
{}from the nucleon rest frame to the pair rest frame.
Introducing the completeness relation for positive energy Dirac 
spinors in Eq.~(\ref{eqn:i}), one finds:
\begin{eqnarray}
I(s_1,s_2,0)&=&\sum_{\bar s_1 \bar s_2}
\overline{u}(\vec k^{{\rm c.m.}}_1,s_1)u_D(\vec k^{{\rm c.m.}}_1,\bar s_1)
\overline u_D(\vec k^{{\rm c.m.}}_1,\bar s_1)\nonumber\\
&&\gamma ^5\, C\, \overline{u}^\top_D(\vec k^{{\rm c.m.}}_2,\bar s_2)
\left(\overline{u}(\vec k^{{\rm c.m.}}_2,s_2)
u_D(\vec k^{{\rm c.m.}}_2,\bar s_2)\right)^\top
\ , \label{eqn:i2}
\end{eqnarray}
{}from which the Clebsch-Gordan coefficients appear by using
the Dirac spinors in Eq.(\ref{eqn:i2})
\begin{eqnarray}
\overline u_D(\vec k^{{\rm c.m.}}_1,\bar s_1)\gamma ^5 C
\overline{u}^\top_D(\vec k^{{\rm c.m.}}_2,\bar s_2)
\rightarrow \chi_{\bar s_1}^\dagger i  \sigma_2\chi^*_{\bar s_2}
=\sqrt{2}\;\langle \frac{1}{2} \bar s_1  \frac{1}{2} \bar s_2|
00\rangle \ . \label{i3}
\end{eqnarray}

The Melosh rotations of the quark spins in the quark-nucleon coupling
are made explicit using Eqs. (\ref{mel}), (\ref{nuc}), (\ref{eqn:i2}),
and (\ref{i3}),
\begin{equation}
\chi (s_1,s_2,s_3;s_N)= \sum_{\bar s_1 \bar s_2}
\left[ R^\dagger_M(\vec k_1^{{\rm c.m.}})\right]_{s_1\bar s_1}
\left[ R^\dagger_M(\vec k_2^{{\rm c.m.}})\right]_{s_2\bar s_2}
\left[ R^\dagger_M(\vec k_3)\right]_{s_3 s_N}
\chi_{\bar s_1}^\dagger i\sigma_2\chi^*_{\bar s_2} \;
\ , \label{numel}
\end{equation}
where the momentum arguments of the Melosh rotations of the spin-zero
coupled pair (12) in Eq.~(\ref{numel}) are taken in the rest frame of
the pair. For the third particle arguments of the Melosh rotation are
taken in the nucleon rest frame. That differs from the BT construction
where the arguments of the Melosh rotations are all taken in the
nucleon rest frame.  Moreover, the various total momentum $'+'$
components, $p^+_{12}$ and $p^+$ in Eq.(\ref{numel}) now appear in
different frames whereas in the BT case only $M_0$ occurs in place of
$p^+$.

In the nucleon rest frame the pair-spin 0 invariant related to
$\rlap\slash p+m_N$ ($\alpha=1/2$) reduces to the projector
$\gamma_0+1$. This means that also the momentum arguments of the
Melosh rotations are taken in the nucleon rest frame. Note, however,
that this case still differs from the BT construction because the sum
of the $'+'$ components of the quark momenta adds to the nucleon
momentum $p^+$ and not to $M_0$ as in the BT formalism. The difference
between BT and the effective Lagrangian quark spin couplings used
here appears in a vanishing limit of the nucleon radius as the
internal quark transverse momentum diverges while the arguments of the
Melosh rotations obtained through the BT construction or the effective
Lagrangian are distinct.  In particular, the nucleon weak isovector
axial vector coupling constant shows a peculiar behavior in the limit
of a pointlike nucleon.

To give a more explicit example we recall the expression of the axial
vector coupling constant found in the context of the BT construction
\cite{bsch,salme}
\begin{eqnarray}
g^{BT}_A=\frac53 \langle \  \frac{(m+x_3 M_0)^2- k^2_{3\perp}} 
{(m+x_3 M_0)^2+ k^2_{3\perp}}\  \rangle \ ,
\label{gabt}
\end{eqnarray} 
where the expectation value is evaluated with the square of the
momentum part of the wave function; $x_3$ is the light-front momentum
fraction with values  bounded by $0 <x_3<1$.  The prescription
given by the effective Lagrangian roughly amounts to substituting the
free three quark mass $M_0$ by the nucleon total $p^+$ which is
$m_N$ in this case, viz.
\begin{eqnarray}
g_A\approx\frac53 \langle \ \frac{(m+x_3 m_N)^2- k^2_{3\perp}} 
{(m+x_3 m_N)^2+ k^2_{3\perp}} \  \rangle \ .
\label{gael}
\end{eqnarray} 
In the limit of a pointlike nucleon ($\beta \to \infty$ is the zero
radius limit corresponding to the strong relativistic limit, i.e.,
$|\vec k_{3\perp}| \gg m+x_3 m_N $) the operator in Eq.(\ref{gael})
tends to $-1$, while in Eq.(\ref{gabt}) the term that contains the
free mass cannot be neglected. From the evaluation of Eq.(\ref{gael})
in this limit one obtains $g_A\approx -5/3$ a value that is
approximately found in our calculations. The pointlike nucleon limit
is a scale invariant point in the sense that the other sensible
physical scales, i.e. nucleon and quark masses, are irrelevant for the
physics. This idea has its origin in the scale invariance of $g_A$ in
quark confining potential models \cite{tegen}, however we stress that
in our case only one situation has this property of scale invariance,
i.e.  the limit of $\beta\rightarrow \infty$. In the next section the
numerical results of the electroweak nucleon properties are shown for
different momentum parts of the wave function as well as for
different quark spin couplings to the nucleon as given by the
effective Lagrangian Eq.(\ref{lag}).

\section{Results and Discussion}

In this section we show the effects of different relativistic spin
couplings and momentum wave functions of constituent quarks for
nucleon electroweak properties. The correlations between the static
electroweak observables are investigated with a different momentum
part of the nucleon light-front wave function for each quark spin
coupling scheme. The Fock state component of the nucleon corresponding
to three constituent quarks as the main part is a strong constraint on
the static observables, and the results are mostly dependent on the
constituent quark mass and one more static observable. Among the
observables the neutron charge radius plays a special role; its
correlation with the magnetic moment dependents on the quark spin
coupling scheme. The parameters of the model are given in Table I.

To discuss the neutron charge radius in some detail we define an
auxiliary dimensionless function $\xi(q^2)$,
\begin{equation}
F_{1n}(q^2)\equiv \frac{q^2}{4m_N^2} \;\xi(q^2)
\end{equation}
that simply reparameterizes the neutron Dirac form factor
$F_{1n}(q^2)$. Since $F_{1n}(0)=0$ the function $\xi(q^2)$ serves as a
``magnifying glass'' for the region $q^2\simeq 0$. In turn
\begin{equation}
G_{En}(q^2)=\frac{q^2}{4m_N^2} \;
(G_{Mn}(q^2)+\xi(q^2))+O(q^4).
\end{equation}
The charge radius is then
\begin{equation}
\langle r^2_{En}\rangle
=\frac{3}{2m_N^2}\;(\hat\mu_n + \xi(0))
\end{equation}
where the neutron magnetic  moment is given by
$\mu_n=\hat\mu_n\,\mu_N$. Using the experimental value~\cite{kop}
 for $\langle
r^2_{En}\rangle$ we find
\begin{equation}
\xi^{\mathrm{expt.}}(0)=0.21\pm 0.08.
\end{equation}
An interesting question is related to a possible restriction of the
values of $\xi(0)$. Presently, the well known Foldy approach to the
charge radius is achieved by
\begin{equation}
\xi^{\mathrm{Foldy}}(0)=0
\end{equation}
that leads to $\langle
r^2_{En}\rangle^{\mathrm{Foldy}}=-0.126 \mathrm{~fm}^2$.
For the naive SU(6) quark model $\langle
r^2_{En}\rangle^{\mathrm{SU(6)}}=0$ that is achieved by
\begin{equation}
\xi^{\mathrm{SU(6)}}(0)=-\hat\mu_n.
\end{equation}
Our model results for $\xi(0)$ obtained with the parameters of Table I
are shown in Table II.

In Table III, we compare our calculations with those of Konen and
Weber\cite{kw} using a Gaussian wave function with the width parameter
$\beta$ that fits $g_A$ using quark masses $m$ of 330, 360, and 380
MeV. Their calculations have the spinors of the pair projected on the
upper components in the nucleon rest-frame and correspond exactly to
the choice $\alpha=1/2$. Our results are in agreement with those
obtained in Ref.\cite{kw}. For each $m$ and $\beta$, we show results
for with $\alpha= 1$ and 0.  This shows that the effect of the
modified quark-pair rest-frame Melosh rotations discussed above are
important and evidenced through the dependence on $\alpha$ which is
also noticeable in the sign of the neutron square radius, as discussed
already in \cite{asfbw}.

\subsection{Static Observables}
{}From now on we use a quark mass of 220 MeV that has been widely used
in connection with realistic models for the meson and nucleon
phenomenology \cite{salme}. In Figs. 2 to 7 we show results for the
correlations between static nucleon electroweak properties, viz.
neutron charge radius, proton radius, magnetic moments and weak
isovector axial vector coupling. Our calculations are done for
different spin couplings of quarks, i.e.  $\alpha=0$, 1/2, 1 in the
effective Lagrangian of Eq.(\ref{lag}), and momentum wave functions of
a harmonic oscillator (HO) (Gaussian) and a power-law (Power) form
($p=3$), viz.  \setcounter{eqntmp}{\value{equation}}
\setcounter{equation}{\value{eqnaux}}
\begin{equation}
\Psi_{\mathrm{HO}}=N_{\mathrm{HO}}\exp(-M^2_0/2\beta_G^2), \qquad
\Psi_{\mathrm{Power}}=N_{\mathrm{Power}}(1+M^2_0/\beta_P^2)^{-p} \ .
\end{equation}
\setcounter{equation}{\value{eqntmp}}

The correlation of the static observables is given by varying the
$\beta$ parameter. Two limits are noteworthy, $\beta \rightarrow 0$
that leads to an infinite size of the nucleon corresponding to the
nonrelativistic limit and $\beta\rightarrow \infty$ that is the zero
radius limit corresponding to the strong relativistic limit.

In Figure 2 results are shown for the neutron charge radius as a
function of the neutron magnetic moment for $\alpha=0$, 1/2, and 1 as
well as HO and Power momentum wave functions.  The results are quite
insensitive to the different shapes of the momentum wave functions,
however strongly dependent on the quark spin coupling.  The neutron
charge radius is a result of a delicate cancellation between
the different contributions to the current in Eq.(\ref{mjp})
and therefore it is strongly sensitive to different quark spin
couplings\cite{asfbw}. Here we extend the conclusion of our previous
work \cite{asfbw}, namely, the neutron charge radius favors the scalar
coupling between the quark-pair also for different forms of momentum
wave functions. The gradient spin coupling ($\alpha=0$) is again found
in complete disagreement with the experimental data. This conclusion
is further supported by the results of the neutron charge form factor
shown later in Figure 8.

The correlation between the magnetic moments of the nucleons is shown
in Figure 3.  The different models of quark spin couplings (for
$\alpha$ equal to 0, 1/2 and 1) in the plot of $\mu_p$ against $\mu_n$
represent a systematic pattern that is again quite independent of the
shape of the momentum wave function. For the chosen constituent mass
$m=220$ MeV the data are not reproduced. The scalar coupling has a
stronger discrepancy than the gradient coupling.  For the scalar case
a change of the constituent mass to about 1/3 of the nucleon mass
still does not lead to a satisfactory result. For $\beta$ going to
infinity the model represents a pointlike particle with the nucleon
anomalous magnetic moments tending towards zero. This limit although
not shown in the figure is achieved in our calculations that explains
the decreasing behavior of $\mu_p$ as a function of $\mu_n$.

The functional dependence of the proton magnetic moment on the
dimensionless product of nucleon mass and proton charge radius ($m_N
r_p$) is shown in Figure 4. We basically reproduce the results
previously found within the Bakamjian-Thomas spin coupling
scheme\cite{bsch}.  We note that Ref.\cite{bsch} used a proton radius
given by the slope of the Dirac form factor $F_1(Q^2)$. For the
different spin coupling schemes there is a weak dependence of $\mu_p$
on the shape of the momentum wave function and moreover the dependence
on different $\alpha$'s is small.

The weak isovector axial vector coupling constant $g_A$ as a function
of the neutron magnetic moment is shown in Figure 5.  Our calculations
for $\alpha=1/2$ and harmonic oscillator wave function are in complete
agreement with those of Konen and Weber\cite{kw}, see Table III.  The
dependence on the shape of the momentum wave function is weak while
increasing the constituent mass would allow us to achieve an agreement
of the scalar quark coupling and the experimental data.  The effective
Lagrangian for the quark-nucleon coupling leads to an axial vector
coupling constant that changes sign in the limit of a pointlike
nucleon. This feature is not present in the Bakamjian-Thomas
construction\cite{bsch} as discussed in the previous section. In the
limit $\beta\rightarrow 0$ the results for $g_A$ tend to the
nonrelativistic value of 5/3 and in the limit of $\beta\rightarrow
\infty$ corresponding to $\mu_n\rightarrow 0$ the axial coupling $g_A$
tends to $\approx -5/3$.

While the change in $\alpha$ has a considerable effect on $g_A$ for a
given neutron magnetic moment (see Fig. 5) this behavior is not seen
for $g_A$ as a function of the proton magnetic moment shown in Figure
6. The momentum shape of the wave function and different values of
$\alpha$ produce small effects on the function $g_A(\mu_p)$. Only the
constituent mass can considerably shift the curve and from Table III
we conclude that the experimental point can be reached with a mass of
about 1/3 of the nucleon mass. However, the simultaneous fit of
$\mu_p$, $\mu_n$ and $g_A$ for $\alpha=1$ seems difficult without
invoking further physical aspects of the constituent quarks.

In Figure 7 the function defined by $g_A(m_nr_p)$ has a weak dependence
on momentum wave function form and spin coupling schemes. This result
could be anticipated from the strong correlations of $g_A(\mu_p)$ 
 and $\mu_p(m_nr_p)$ shown in Figures 6 and 4, respectively. The
experimental point could be fitted by the increase of the constituent
mass. 

{}From the results shown in Figures 2 to 7 we conclude that without
invoking more physics than is contained in the present model, each set
of static observables either $\{r_n, r_p, \mu_p, g_A\}$ or $\{r_n,
\mu_n, g_A\}$ can be reasonably fitted to the experimental values with
only two parameters, i.e. the width of the wave function and the
constituent quark mass. The difficulty is related to the precise and
simultaneous fit of the magnetic moments as shown in Figure 3.

\subsection{Nucleon Form Factors} 

In Figures 8 to 13 we show different electromagnetic and weak form
factors as a function of $q^2$. We give results with the
parameters of the Gaussian and power-law wave functions as given in
Table I. For each $\alpha$ they are fitted to the neutron magnetic
moment.

The neutron charge form factor is shown in Figure 8.  The gradient
spin coupling gives a negative contribution for $-q^2\lesssim 2$
(GeV/c)$^2$.  The calculation for the mixed case ($\alpha=1/2$)
underestimates the data.  For the scalar quark spin coupling both
types of momentum wave functions give results close to each other and within
the experimental uncertainty agree with the data. For momentum
transfers above 1 GeV/c, the model dependence (Power vs. HO) starts to
appear in the neutron charge form factor.

The theoretical results for $G_{Mn}(q^2)$ are compared to the
experiments in Figure 9. The calculations with scalar coupling between
the quark pair ($\alpha=1$) give the best agreement with the data for
both momentum wave function models. The results for $\alpha=0$ and 1/2
overestimate the data.  For $ -q^2\gtrsim 1$ (GeV/c)$^2$ the models
deviate from experiments.

In Figure 10 we show the proton charge form factor compared with
experiments.  A common behavior is found for the calculations with
both wave function models, i.e., the choice of $\alpha=1$ gives values
below the experimental data.  This could also be anticipated from
Figures 3 and 4 that show too big values of the proton radius for
$\mu_n=-1.91\mu_N$.  The spin couplings given by $\alpha=0$ and 1/2
approach the data for $-q^2\lesssim  2$ (GeV/c)$^2$,
because the proton radius is in better agreement with the experimental
values. 

In Figure 11 the results for the proton magnetic form factor are
shown. The scalar quark spin coupling results approach experimental
data for momentum transfers below 1 GeV/c and for both wave function
models.  The results obtained with the spin coupling parameterized by
$\alpha=0$ and 1/2 overestimate the data.

In Figure 12 the results of recent measurements of the ratio
$\mu_pG_{Ep}/G_{Mn}$ \cite{jones} are compared to our calculations. We
observe a dependence on different spin couplings and momentum wave
functions. However the data are generally underestimated that
indicates the necessity for more sophisticated wave function models,
inclusion of other spin couplings, and/or a constituent quark
substructure. Let us emphasize that relativistic effects are crucial
for the steeper proton charge form factor fall-off.

Finally, in Figure 13 the model results are compared to experimental
data for the nucleon weak isovector axial vector form factor.  The
calculations with the scalar coupling between the quark pair produce
the best agreement with the data. However a remarkable sensitivity to
the coupling schemes and wave functions models is also seen in Figure
13. The model dependence found in this figure can be qualitatively
understood if one looks at the approximate equation (\ref{gael}) for
$g_A$, where a cancellation between two terms occurs that causes a
high sensitivity to details of the models. This could also be
expected for $q^2$-dependence of the axial vector form factor.  

We must keep in mind that our wave function models are quite
simplistic and even in the nonrelativistic quark model the nucleon is
highly relativistic and the real wave function can strongly differ
from their nonrelativistic counterparts.  In this sense, the
difference between the data and the present models seen in Figures 9
to 13 for momentum transfers of several GeV/c is not too serious
considering the simplicity of the model. We should also mention that
the concept of {\em constituent} quarks is expected to break down
above the chiral symmetry breaking scale ($ 4\pi f_\pi \sim 1$ GeV),
so that we expect the model to loose validity because current quarks
become the relevant degrees of freedom revealing the {\em constituent}
substructure.

\section{Summary and Conclusion}

We have shown the effects of different forms of relativistic spin
couplings of constituent quarks on the nucleon electroweak properties.
Model independent (i.e. independent of the momentum shape of the
light-front wave function) relations between the static electroweak
observables are verified to hold within each quark spin coupling
scheme as could be expected as $q^2\rightarrow 0$. It is found that,
while the neutron charge form factor is very sensitive to different
choices of spin coupling schemes, it is insensitive to the details of
the momentum part of the three-quark wave function model for momentum
transfers below 1 GeV/c. The experimental data on the neutron charge
form factor -- for momentum transfers below 1 GeV/c -- can be
reproduced by models with a scalar coupling of the constituent quark
pair, independent of the shape of the wave function.  This is mostly
due to the momentum dependence in lower component of the quark spinors
that leads to a mixed-symmetry space part (in a nonrelativistic
reduction), compare also Ref.~\cite{Cardarelli:2000tk}. This feature
is strongly suppressed in the mixed case ($\alpha=1/2$) and comes with
an opposite sign for the pure scalar and pure gradient cases,
respectively. 

The difference between Bakamjian-Thomas and effective Lagrangian spin
coupling schemes is particularly noticeable in the weak isovector
axial vector coupling constant evaluated in the pointlike nucleon
limit. The correlations involving the set of static observables
$\{r_p, \mu_p, g_A\}$ are not very sensitive to spin coupling schemes
defined by the effective Lagrangian for different values of $\alpha$
in Eq.(\ref{lag}).  Among these relations, the function $g_A(\mu_p)$
is shown to have the smallest dependence on spin coupling schemes and
on the shape of the momentum wave function. The correlations involving
the neutron magnetic moment are more sensitive to different spin
coupling schemes. Overall, for momentum transfers above 1 GeV/c, we
observe a dependence on the different spin coupling schemes and
momentum wave functions. The new data on the ratio of
$\mu_pG_{Ep}/G_{Mp}$ indicates the necessity to improve the wave
function models, include other (e.g. axial-vector quark pair) spin
coupling, and/or a description of constituent quarks beyond the models
discussed in the present work. The influence of pionic corrections in
a light front frame work has been studied in
~\cite{Dziembowski:1997vh}. From their results we expect that our
conclusions do not change drastically, however, a complete study of
pionic corrections in the present framework is still an open and
challenging problem.

{\bf Acknowledgments:} MB thanks R. Tegen for a discussion on
$g_A$. HJW and MB thank the University of Virginia's INPP for partial
support. MB thanks the Deutscher Akademischer Austauschdienst (DAAD)
and FAPESP for support, and the Department of Physics of ITA for the
warm hospitality and for local support. WRBA thanks CNPq for financial
support and LCCA/USP for providing computational facilities, EFS
thanks FAPESP for financial support and TF thanks CNPq and FAPESP.

\begin{table}
\caption{ Parameters for  the HO  $(\beta_G)$ and 
power-law $(\beta_P)$ models of the nucleon momentum wave function 
with different spin coupling schemes from the fit of $\mu_n=-1.91 
\mu_{\rm N}$ with $m=$ 220 MeV.}
\vspace{0.5 cm}
\begin{tabular}{|c|c|c|}
$\alpha$ &$\beta_G$[MeV]&$\beta_P$ [MeV]\\ \hline
1       &562  &477 \\
1/2     &664  &576  \\
0       &661  &411 
\end{tabular}

\vspace{1cm}
\caption{Values for  $\xi(0)$  from the different models 
(HO, Power) using the parameters of Table I,
$\xi^{\mathrm{expt.}}=0.21\pm 0.08$.}
\vspace{0.5 cm}
\begin{tabular}{|c|c|c|}
$\alpha$&$\xi_G (0)$  & $\xi_P (0)$ \\ \hline
1       & 0.54  & 0.69  \\
1/2     & 1.6  & 1.6   \\
0       & 3.0   & 2.6 
\end{tabular}

\vspace{1cm}
\caption{ Nucleon low-energy electroweak observables 
for different spin coupling parameters with a
gaussian light-front wave function  for 
$m$=330, 360 and 380 MeV with the values of $\beta $ parameter
from  Konen and Weber\protect\cite{kw} 
(in their work the Gaussian parameter is $\beta / \sqrt{3}$ ). }
\vspace{1 cm}
\begin{tabular}{|c|c|c|c|c|c|c|}
$m$[MeV]&$\alpha$ & $r^2_{En}$ [{\rm fm}$^2$] 
& $r^2_{Ep}$ [{\rm fm}$^2$] &
$\mu_n [\mu_{\rm N}]$ &$\mu_p[\mu_{\rm N}]$& $g_A$\\ 
\hline
    & 0   &  0.035 &  0.69 & -1.83 & 2.84 & 1.09\\
330 & 1/2 & -0.024 &  0.69 & -1.73 & 2.80 & 1.20\\
    & 1   & -0.080 &  0.71 & -1.60 & 2.71 & 1.25\\ 
\hline
    & 0   &  0.023 &  0.66 & -1.77 & 2.77 & 1.13\\
360 & 1/2 & -0.025 &  0.66 & -1.67 & 2.72 & 1.23\\
    & 1   & -0.073 &  0.67 & -1.53 & 2.62 & 1.29\\ 
\hline
    & 0   &  0.018 &  0.62 & -1.71 & 2.72 & 1.19\\
380 & 1/2 & -0.027 &  0.62 & -1.61 & 2.66 & 1.20\\
    & 1   & -0.071 &  0.63 & -1.47 & 2.56 & 1.29\\ \hline
&   &  & 0.66 $\pm$ 0.06\cite{brod},&  & &  \\
EXP. & & -0.113 $\pm$ 0.005\cite{kop}&0.74 $\pm$ 0.02\cite{mur},&-1.91 &2.79 & 
1.2670 $\pm$ 0.0035 \cite{pdg2000}  \\
& & &0.77 $\pm$ 0.03\cite{rosen}& & &  
\end{tabular}

\vspace{2 cm}

\end{table}
\begin{figure}[h]
\begin{center}
\vspace{5cm}
\centerline{\
\begin{picture}(330,130)(-5,-130)
\GOval(30,25)(15,5)(0){.5}
\GOval(120,25)(15,5)(0){.5}
\Vertex(80,70){4.0}
\put(110,12){\makebox(0,0)[br]{$(2)$}}
\put(110,42){\makebox(0,0)[br]{$(1)$}}
\put(110,72){\makebox(0,0)[br]{$(3)$}}
\Line(30,10)(120,10)
\Line(30,40)(120,40)
\Line(30,70)(120,70)
\put(85,-5){\makebox(0,0)[br]{$(1a)$}}
\GOval(150,25)(15,5)(0){.5}
\GOval(240,55)(15,5)(0){.5}
\Vertex(200,70){4.0}
\Line(150,10)(240,10)
\Line(150,40)(240,40)
\Line(150,70)(240,70)
\put(210,-5){\makebox(0,0)[br]{$(1b)$}}
\GOval(30,-75)(30,5)(0){.5}
\GOval(120,-60)(15,5)(0){.5}
\Vertex(80,-45){4.0}
\Line(30,-105)(120,-105)
\Line(15,-75)(120,-75)
\Line(30,-45)(120,-45)
\put(85,-120){\makebox(0,0)[br]{$(1c)$}}
\GOval(150,-60)(15,5)(0){.5}
\GOval(240,-60)(15,5)(0){.5}
\Vertex(200,-45){4.0}
\Line(150,-105)(240,-105)
\Line(150,-75)(240,-75)
\Line(150,-45)(240,-45)
\put(210,-120){\makebox(0,0)[br]{$(1d)$}}
\end{picture}
}
\end{center}
\caption{Feynman diagrams for the nucleon electroweak current. The gray blob
represents the spin invariant for the coupled quark pair
in the effective Lagrangian, Eq.(\ref{lag}). The black circle in the
fermion line represents the action of the current operator on the quark.
The current operator can represent either the electromagnetic current or 
the weak isovector axial vector current.
Diagram (1a) represents either $J^+_{aN}$, Eq.(\ref{j+a}), or $A^+_{aN}$, 
Eq.(\ref{a+a}).
Diagram (1b) represents either $J^+_{bN}$, Eq.(\ref{j+b}), or
$A^+_{bN}$, Eq.(\ref{a+b}).
Diagram (1c) represents either $J^+_{cN}$, Eq.(\ref{j+c}),
or $A^+_{cN}$, Eq.(\ref{a+c}).
Diagram (1d) represents $J^+_{dN}$, Eq.(\ref{j+d}). Diagram (1d) does not
contribute to the weak isovector axial vector current due to the 
isoscalar nature of the coupled quark pair.}
\label{fig1}
\end{figure}
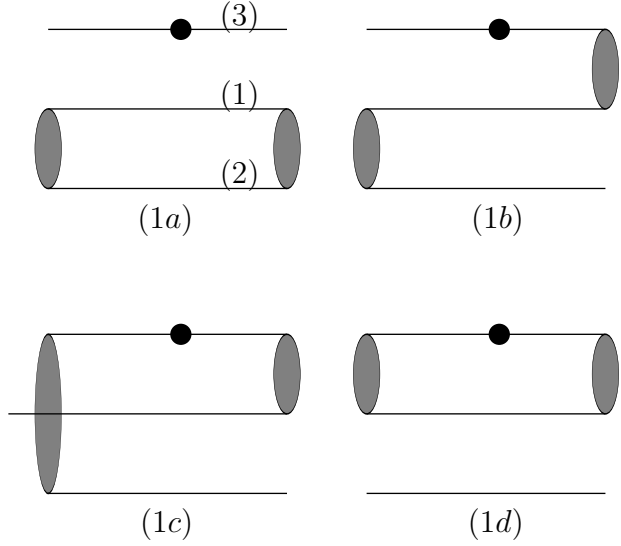
\newpage
 \begin{figure}
 \setlength{\epsfxsize}{0.8\hsize} \centerline{\epsfbox{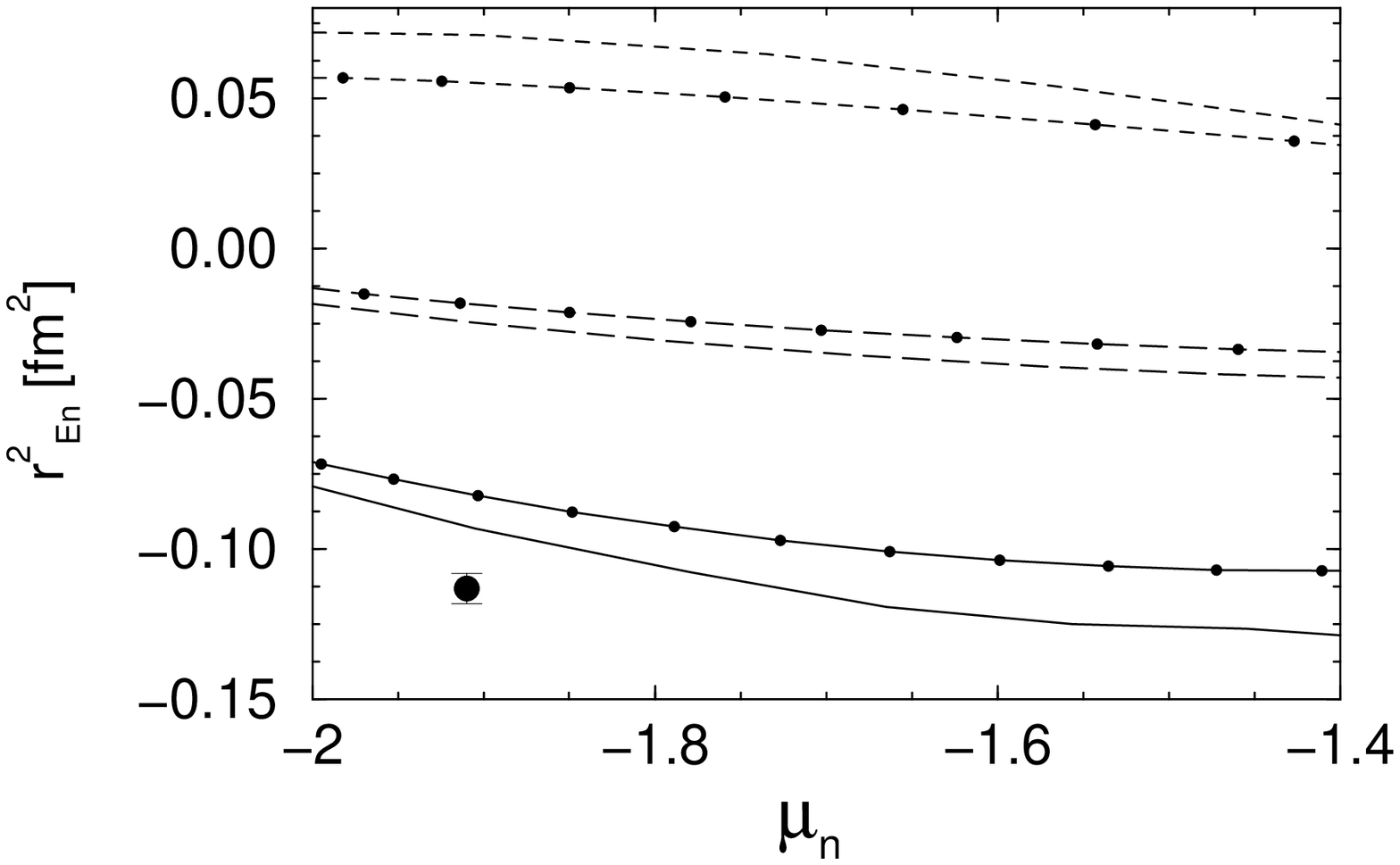}}
 \caption[dummy0]{ Neutron charge square radius as a function of the 
neutron magnetic moment. Results for the Gaussian wave function 
with $\alpha$ equal to 1 (solid line), 1/2 (dashed line) and 0  
(short-dashed line). Results for the power-law wave function 
with $\alpha$ equal to 1 (solid line with dots), 1/2 (dashed line with dots) 
and 0  (short-dashed line with dots). Experimental data from Ref.\cite{kop}.}
\vfill
 \setlength{\epsfxsize}{0.8\hsize} \centerline{\epsfbox{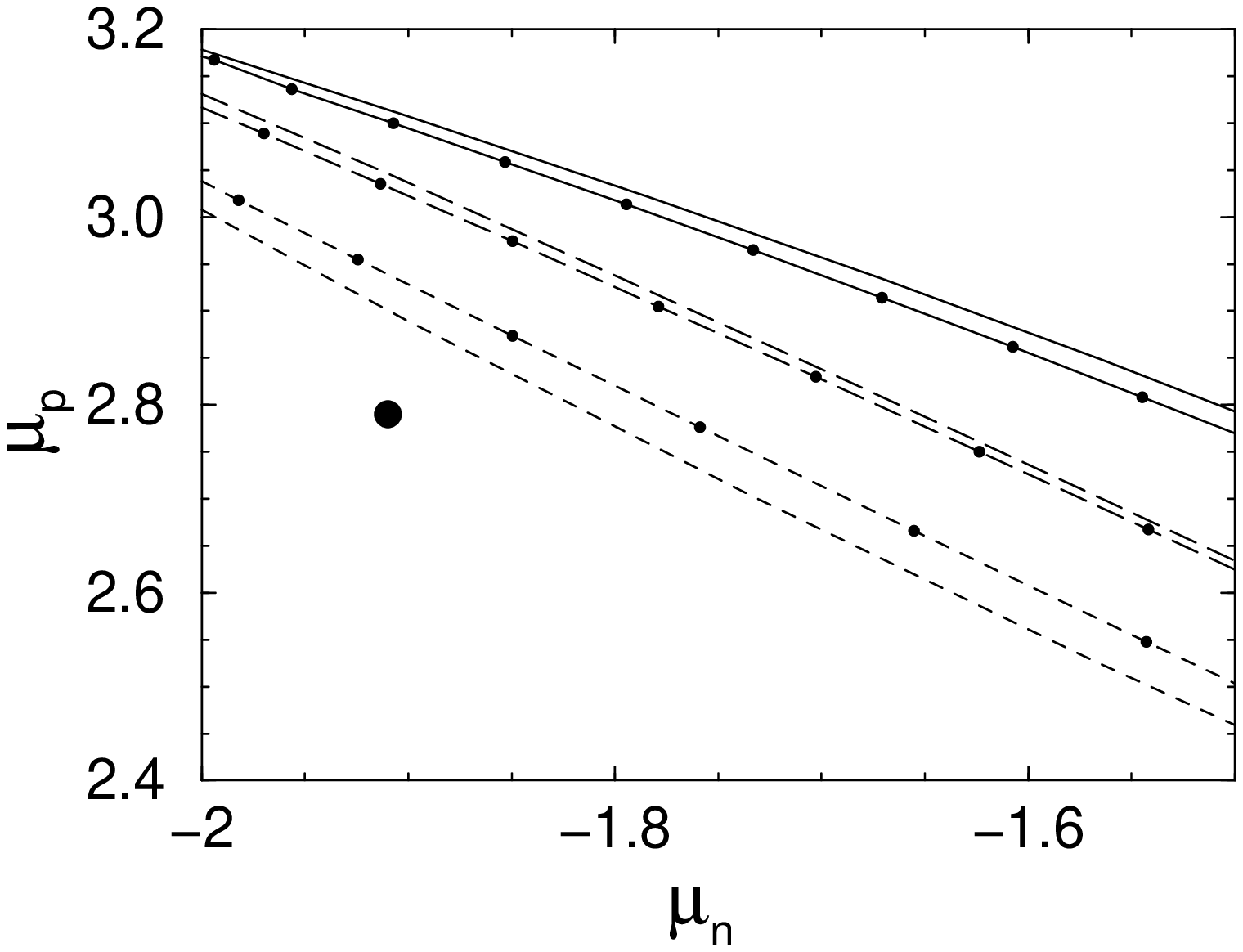}}
 \caption[dummy0]{ Proton magnetic moment as a function 
of the neutron magnetic moment.
Theoretical curves  labeled as in Fig.2. 
The experimental data are represented by the full circle. }
\end{figure}
 \begin{figure}
 \setlength{\epsfxsize}{0.8\hsize} \centerline{\epsfbox{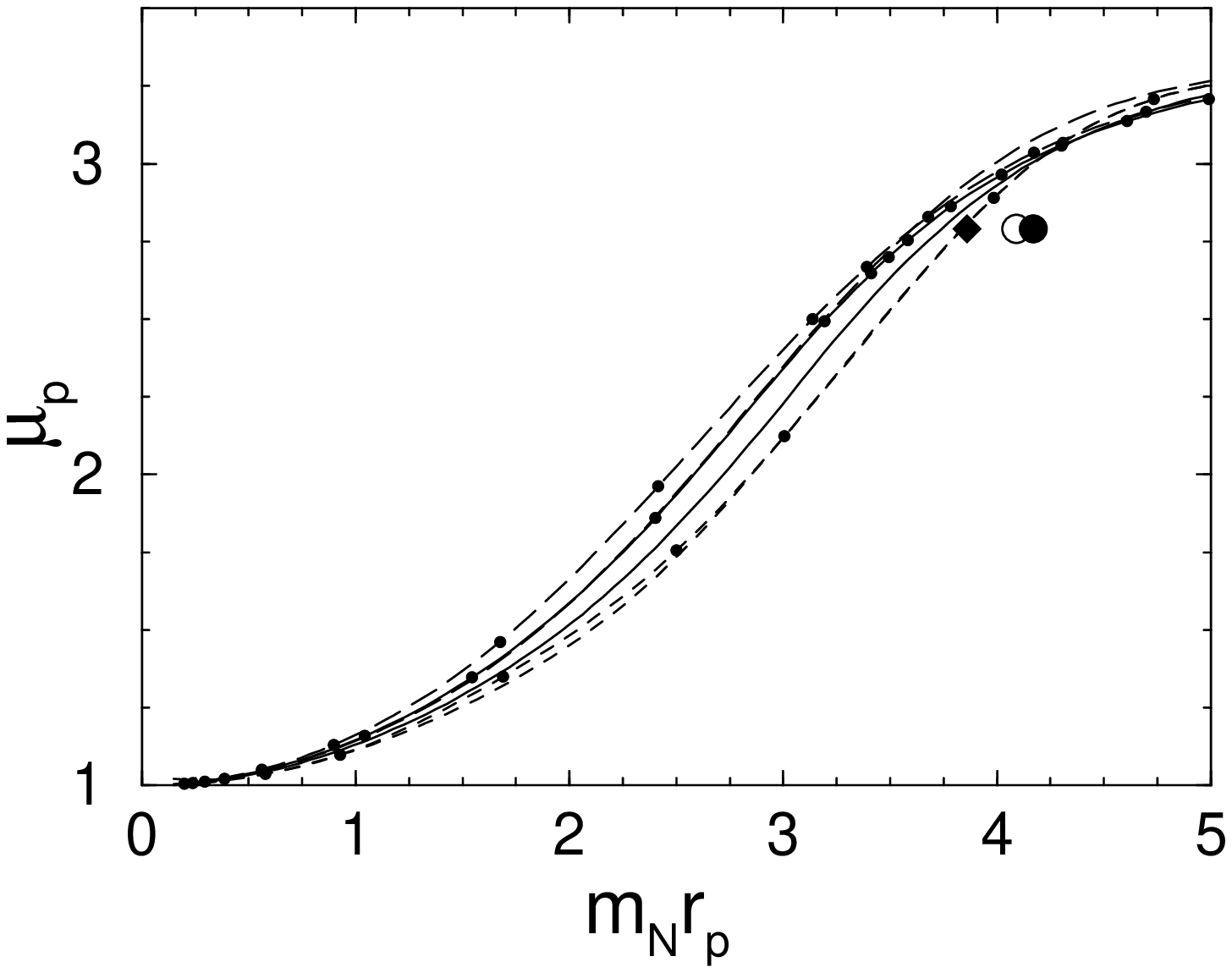}}
 \caption[dummy0]{ Proton magnetic moment as a function 
of the dimensionless product $m_Nr_p$.
Theoretical curves labeled as in Fig.2. Experimental points are
given by a 
full diamond\cite{brod}, open circle\cite{mur} and full circle\cite{rosen}.  }
\vfill
 \setlength{\epsfxsize}{0.8\hsize} \centerline{\epsfbox{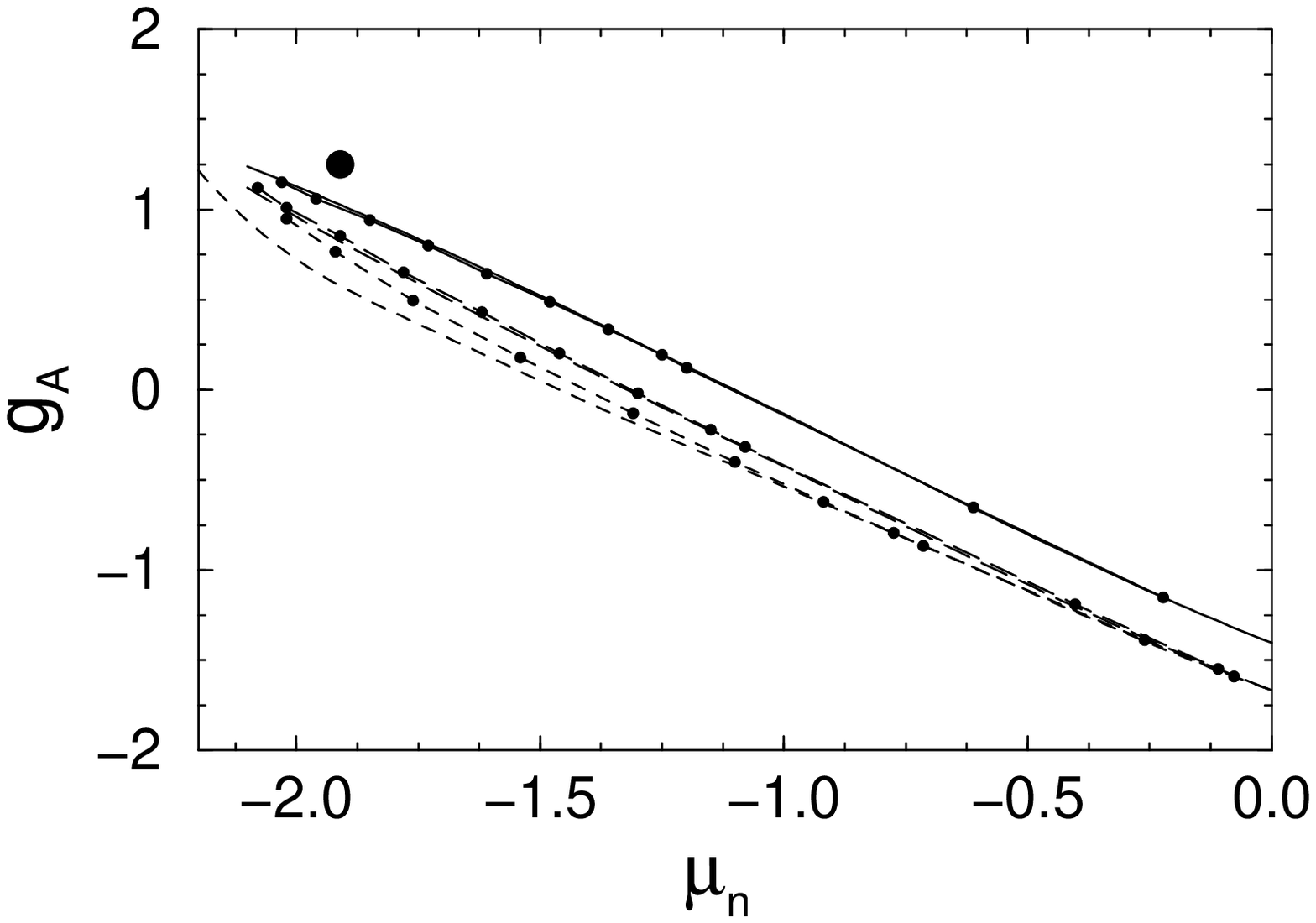}}
 \caption[dummy0]{Nucleon axial vector coupling constant 
as a function of the neutron magnetic moment.
Theoretical curves labeled as in Fig.2. The experimental point
is given by the full circle. }
\end{figure}
 \begin{figure}
 \setlength{\epsfxsize}{0.8\hsize} \centerline{\epsfbox{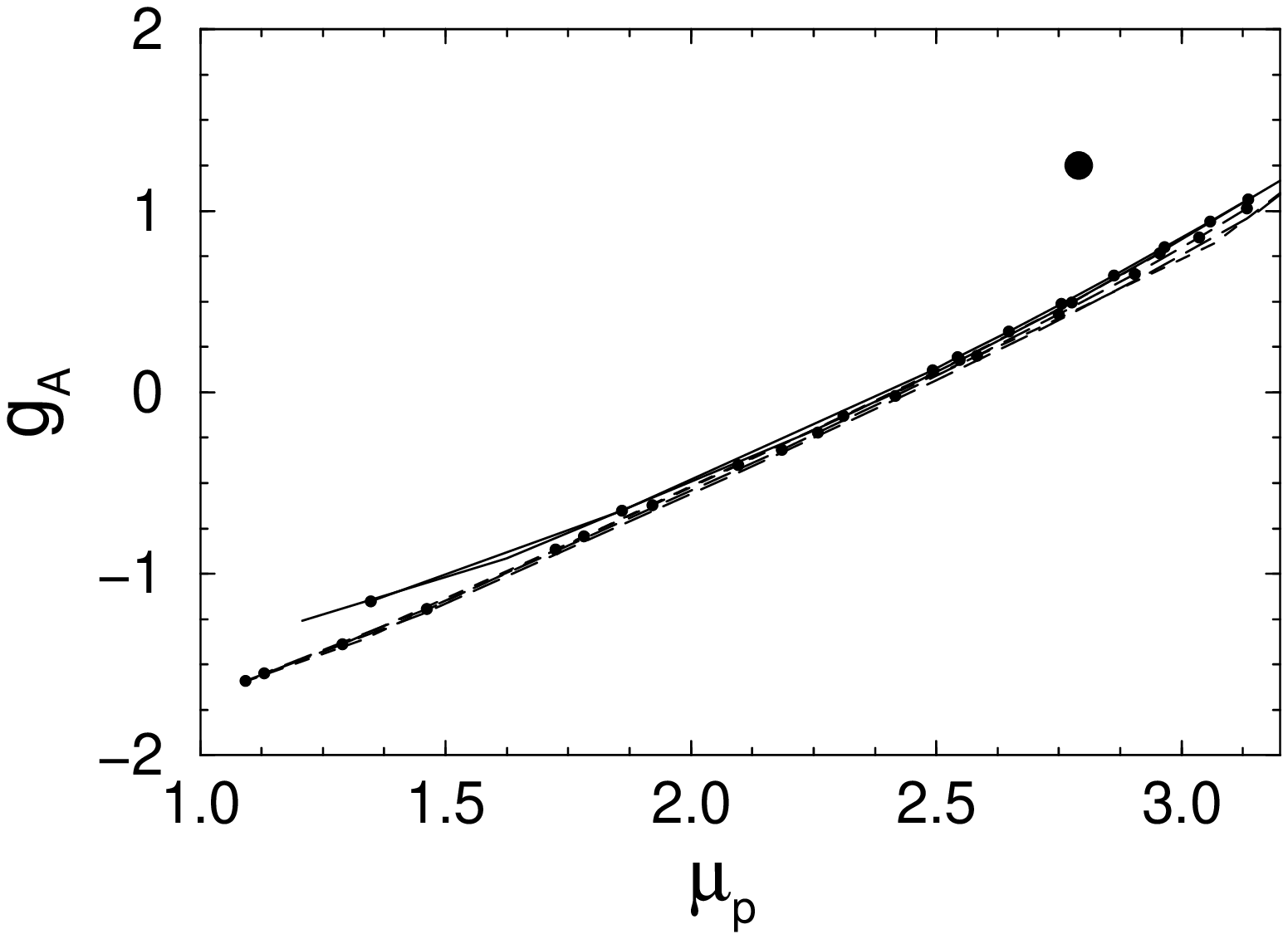}}
 \caption[dummy0]{Nucleon axial vector coupling constant 
as a function of the proton magnetic moment.
Theoretical curves  labeled as in Fig.2. 
The experimental point given by the full circle. }
\vfill
 \setlength{\epsfxsize}{0.8\hsize} \centerline{\epsfbox{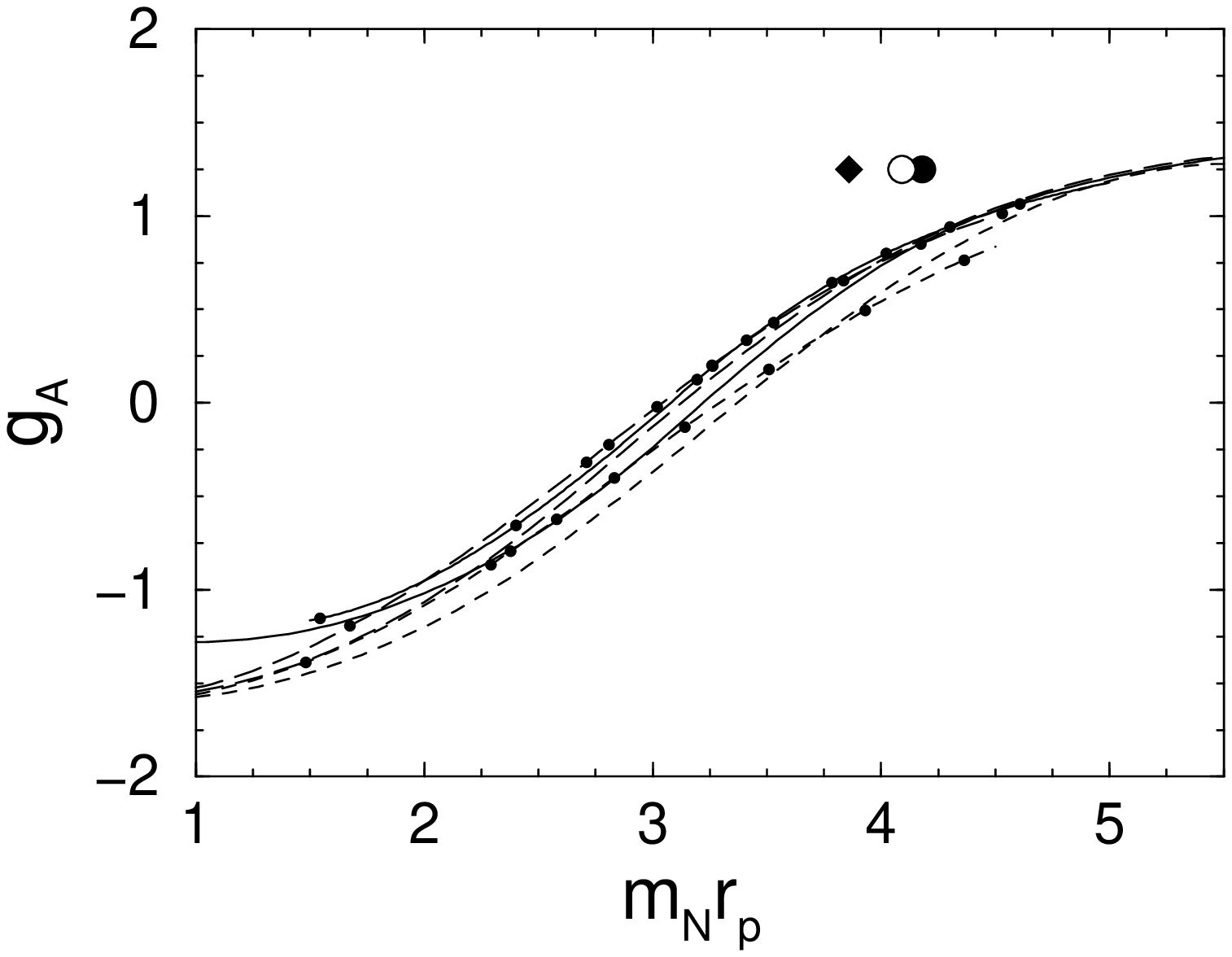}}
 \caption[dummy0]{
 Nucleon axial vector coupling constant 
as a function of the dimensionless product of the proton charge radius and
mass. Theoretical curves  labeled as in Fig.2. 
Experimental points are given by a
full diamond\cite{brod}, open circle\cite{mur} and full circle\cite{rosen}.}
\end{figure}
 \begin{figure}
 \setlength{\epsfxsize}{0.8\hsize} \centerline{\epsfbox{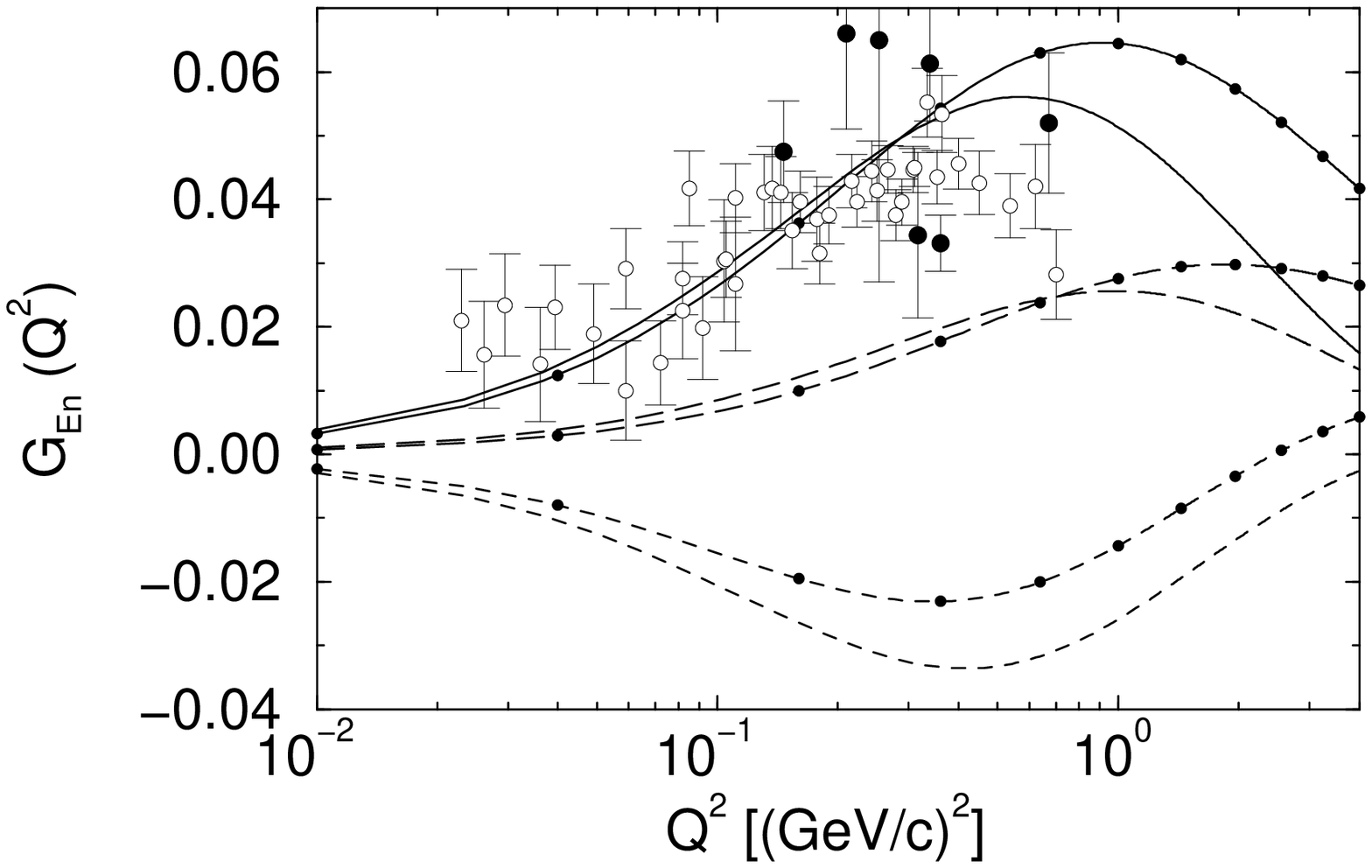}}
 \caption[dummy0]{Neutron charge form factor as a function of the momentum
transfer $q^2=-Q^2$. 
Theoretical curves  labeled as in Fig.2. 
The empty circles are the experimental data from
Ref.\cite{plat} and the full circles from Ref.\cite{eden}.}
\vfill
 \setlength{\epsfxsize}{0.8\hsize} \centerline{\epsfbox{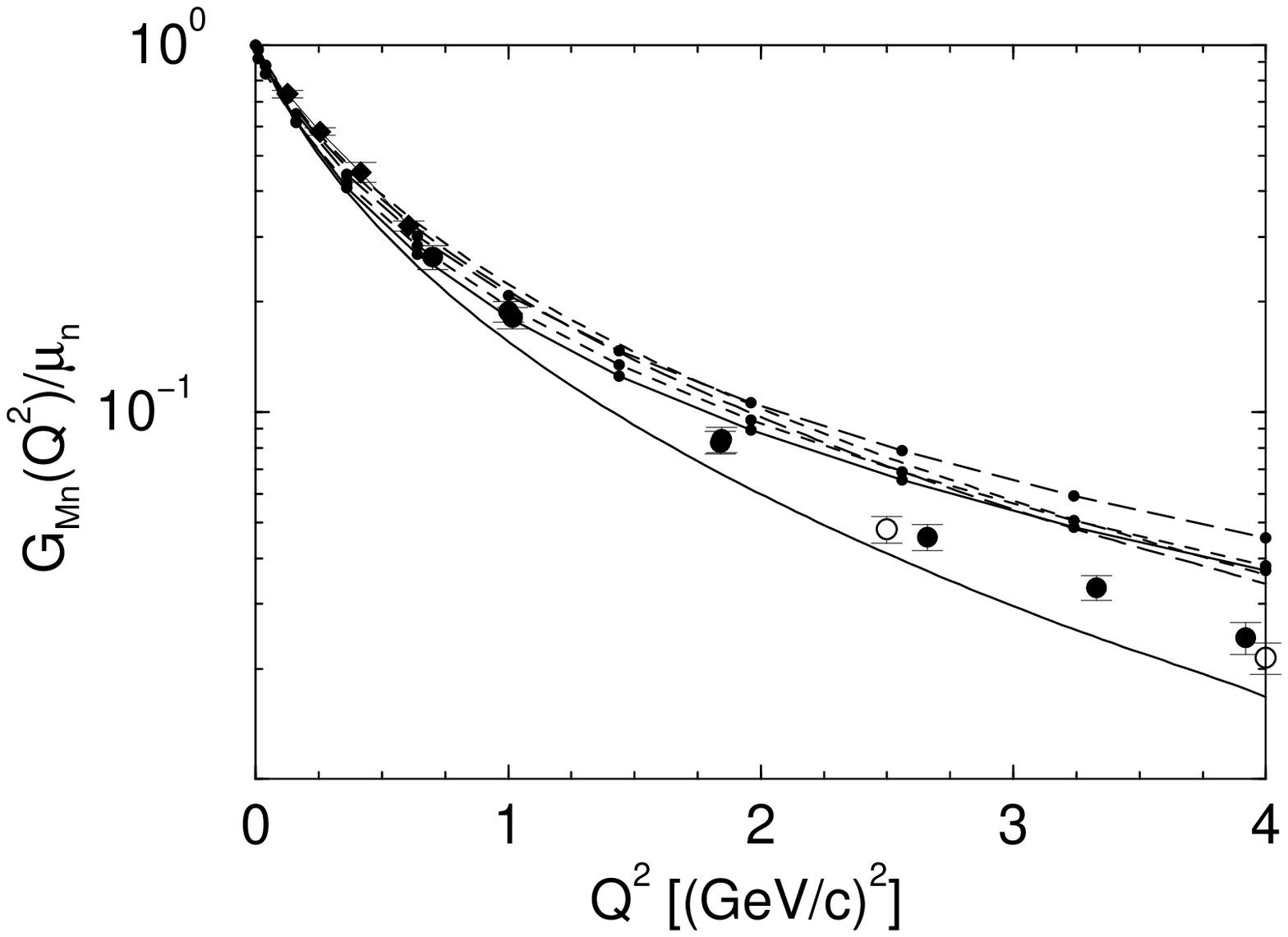}}
 \caption[dummy0]{
Neutron magnetic form factor $G_{Mn}/\mu_n$ as a function
of momentum transfer squared. Theoretical curves labeled as in Fig.2.
The experimental data come from Ref.\cite{alb}, full circles;
Ref.\cite{rock}, open circles; Ref.\cite{bruins}, full diamonds.} 
\end{figure}
 \begin{figure}
 \setlength{\epsfxsize}{0.8\hsize} \centerline{\epsfbox{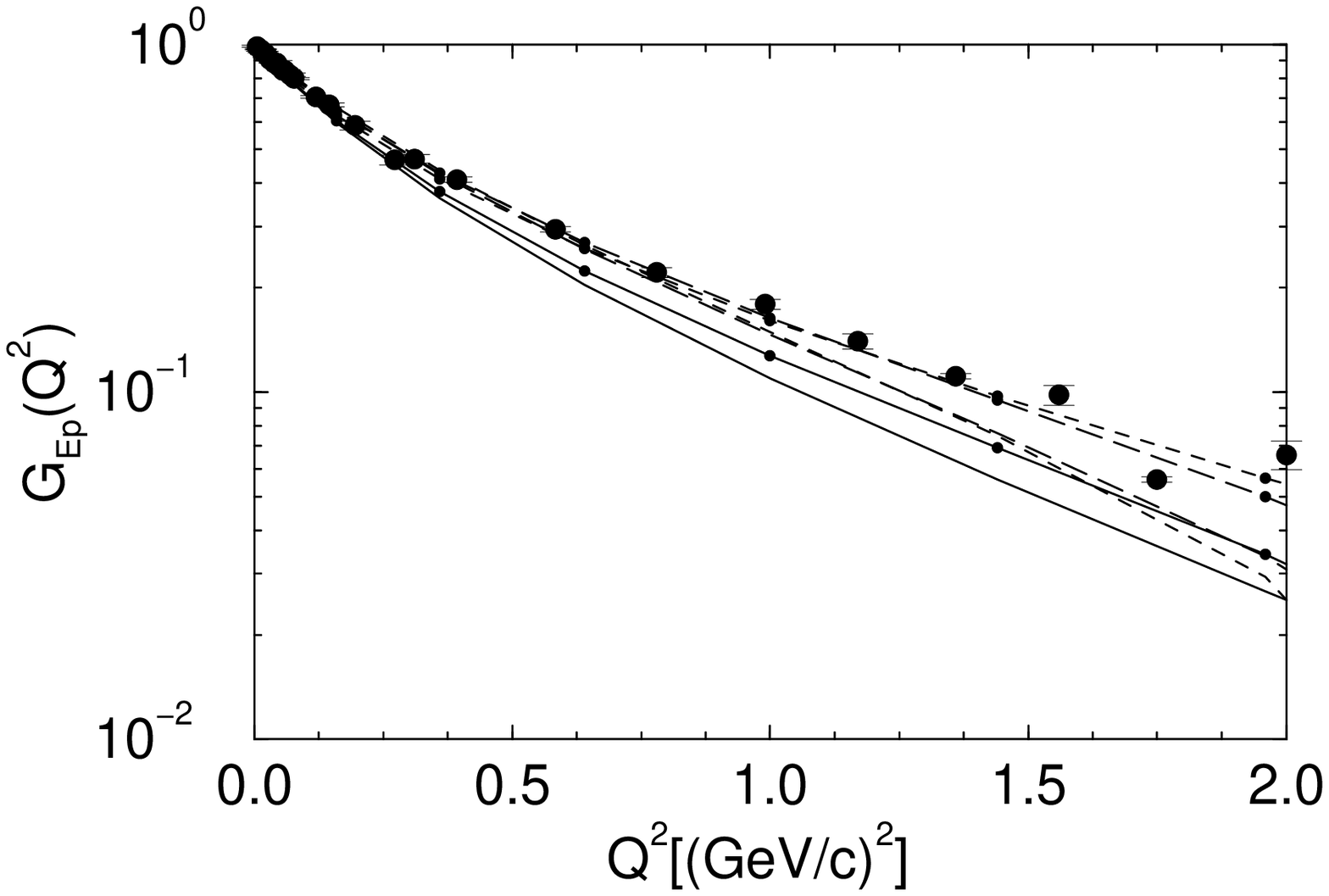}}
 \caption[dummy0]{
Proton charge form factor as a function
of momentum transfer squared. Theoretical curves labeled as in Fig. 2.
The experimental data come from Ref.\cite{hol}.} 
\vfill
 \setlength{\epsfxsize}{0.8\hsize} \centerline{\epsfbox{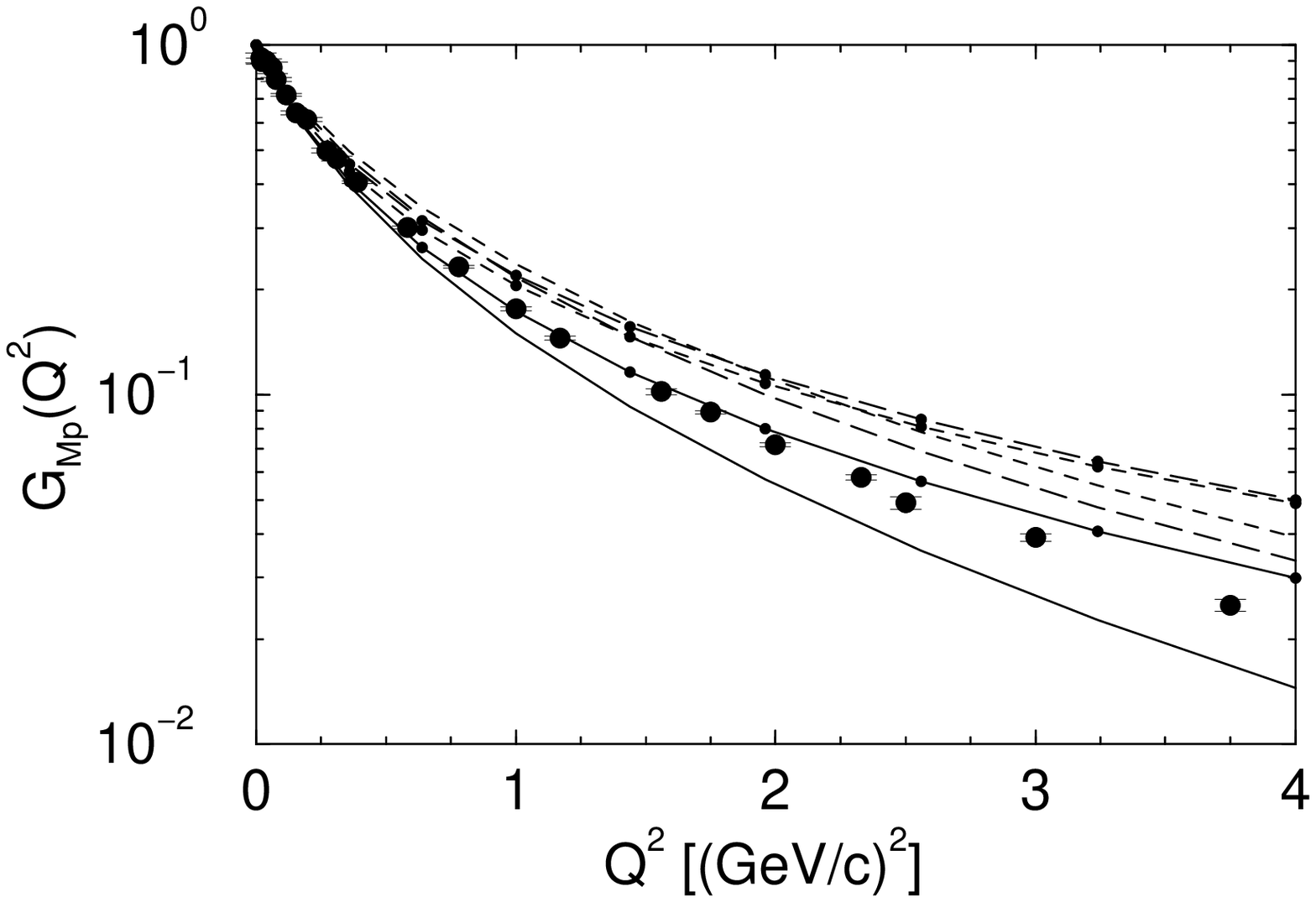}}
 \caption[dummy0]{
Proton magnetic form factor $G_{Mp}/\mu_p$ as a function
of momentum transfer squared. Theoretical curves labeled as in Fig. 2.
The experimental data come from Ref.\cite{bartel}.
 } 
\end{figure}
 \begin{figure}
 \setlength{\epsfxsize}{0.8\hsize} \centerline{\epsfbox{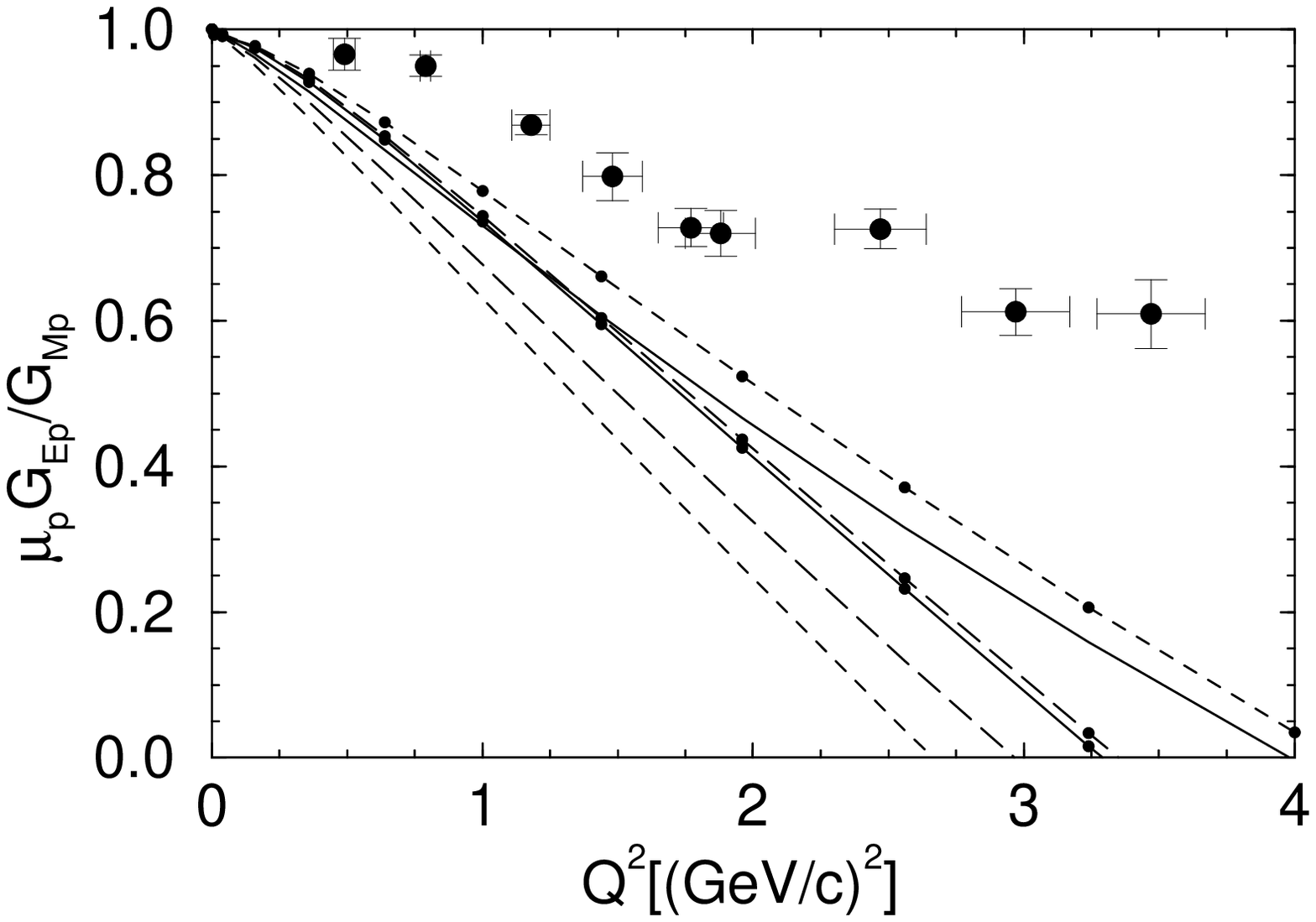}}
 \caption[dummy0]{
Proton form factor ratio $\mu_pG_{Ep}/G_{Mp}$ as a function
of momentum transfer squared. Theoretical curves labeled as in Fig. 2.
The experimental data come from Ref.\cite{jones}.
 } 
\vfill
 \setlength{\epsfxsize}{0.8\hsize} \centerline{\epsfbox{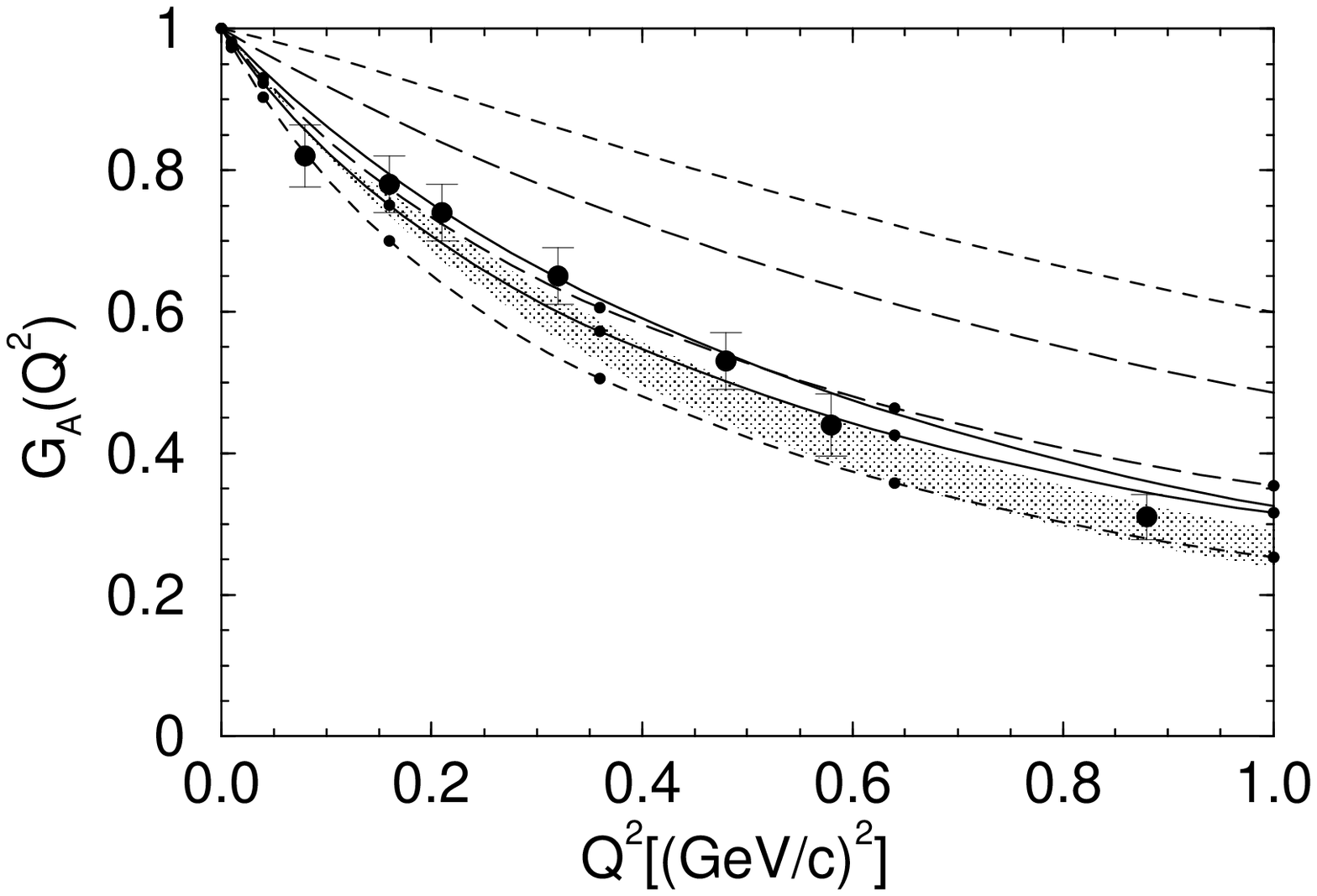}}
\caption[dummy0]{ Normalized axial vector form factor as a
 function of momentum transfer squared. Theoretical curves labeled as
 in Fig. 2.  The experimental data come from Ref.\cite{del}. The
 experimental data of Ref.\cite{exga} are given in terms of a dipole
 form with a combined fit of $m_A=1.03\pm 0.05$ GeV. }
\end{figure}
\end{document}